\documentclass[10pt, final]{IEEEtran}

\usepackage{lipsum}
\usepackage{amsthm}
\usepackage{amsmath}
\usepackage{bbm}

\usepackage{algorithm}
\usepackage{algorithmic}

\usepackage{mathrsfs}
\usepackage{cite}
\usepackage{bm,epsfig,amsthm,url}
\usepackage{indentfirst}
\usepackage{amssymb}
\usepackage{amsfonts}
\usepackage{epstopdf}
\usepackage[caption=false,font=footnotesize]{subfig}
\usepackage{xcolor}
\usepackage{mathtools}

\usepackage{hyperref}
\usepackage{balance} 

\theoremstyle{definition}

\newtheorem{remark}{Remark}

\usepackage{graphicx}
\usepackage{epstopdf}

\usepackage{framed}

\begin{document}

\title{Intelligent Reflecting Surface Aided Multi-Tier Hybrid Computing} 

\author{
Yapeng~Zhao,
Qingqing~Wu,
Guangji~Chen,
Wen~Chen,
Ruiqi~Liu,
Ming-Min~Zhao,
Yuan~Wu,
~and Shaodan~Ma


\thanks{Y. Zhao is with the State Key Laboratory of Internet of Things for Smart City, University of Macau, Macao 999078, China, and also with the Department of Electronic Engineering, Shanghai Jiao Tong University, 200240, China (email: yc17435@connect.um.edu.mo).  	
Q. Wu and W. Chen are with the Department of Electronic Engineering, Shanghai Jiao Tong University, 200240, China (e-mail: qingqingwu@sjtu.edu.cn; wenchen@sjtu.edu.cn). 
G. Chen, S. Ma and Y. Wu are with the State Key Laboratory of Internet of Things for Smart City, University of Macau, Macao 999078, China (email: guangjichen@um.edu.mo; shaodanma@um.edu.mo; yuanwu@um.edu.mo).
R. Liu is with the State Key Laboratory of Mobile Network and Mobile Multimedia Technology, ZTE Corporation, Shenzhen 518057, China (e-mail: richie.leo@zte.com.cn).
M.-M. Zhao is with the College of Information Science and Electronic Engineering, Zhejiang University, Hangzhou 310027, China, and also with the Zhejiang Provincial Key Laboratory of Information Processing, Communication and Networking (IPCAN), Hangzhou 310027, China (e-mail: zmmblack@zju.edu.cn).
}
}

\maketitle

\begin{abstract}
The digital twin edge network (DITEN) aims to integrate mobile edge computing (MEC) and digital twin (DT) to provide real-time system configuration and flexible resource allocation for the sixth-generation network. This paper investigates an intelligent reflecting surface (IRS)-aided multi-tier hybrid computing system that can achieve mutual benefits for DT and MEC in the DITEN. 
For the first time, this paper presents the opportunity to realize the network-wide convergence of DT and MEC.
In the considered system, specifically, over-the-air computation (AirComp) is employed to monitor the status of the DT system, while MEC is performed with the assistance of DT to provide low-latency computing services. Besides, the IRS is utilized to enhance signal transmission and mitigate interference among heterogeneous nodes.
We propose a framework for designing the hybrid computing system, aiming to maximize the sum computation rate under communication and computation resources constraints. To tackle the non-convex optimization problem, alternative optimization and successive convex approximation techniques are leveraged to decouple variables and then transform the problem into a more tractable form.
Simulation results verify the effectiveness of the proposed algorithm and demonstrate the IRS can significantly improve the system performance with appropriate phase shift configurations.  
Moreover, the results indicate that the DT assisted MEC system can precisely achieve the balance between local computing and task offloading since real-time system status can be obtained with the help of DT. 
\end{abstract}

\section{Introduction}

The sixth-generation (6G) network has been conceptualized for several years since the worldwide deployment of the fifth-generation (5G) wireless networks \cite{WalidSaad_6GVision,Towards6G, RuiqiLiu}.
The ultimate goal of 6G is to integrate communication, sensing, computing, and intelligence, leading to the development of the Intelligent Internet of Everything (IIoE) \cite{XiaoyangLi_WC23}.
This integration will be instrumental in providing seamless intelligent services for computation-intensive and delay-sensitive applications, including autonomous driving and robotic factory \cite{YuanmingShi_JSAC22_EdgeAI}. To facilitate the realization of these fancy applications, the 6G network must be able to accommodate the participating devices for real-time computing, communication, and control.
However, such networks typically consist of massive heterogeneous nodes with varying communication, computing capabilities, and functionalities, posing a critical challenge to network operation. For instance, the end user equipment (UE) which requires intelligent services are generally with limited computation-resource due to strict size constraints and production costs \cite{MaoYZHL17}.
As a result, UEs need to offload their workloads to servers that are far away from the generated data.
This results in a complex system, and its effective design becomes a critical problem that requires careful consideration.

The digital twin (DT) technology offers a promising solution for the effective use of available communication and computation resources in 6G, as it can digitize the physical world through real-time interaction with the physical environment. 
DT collects multidimensional data from the physical entities, such as sensor updates and system operating history, to monitor their status and reflect their entire life cycles \cite{CM22_DT_6G}. 
DT was initially coined in 2003 \cite{IoTJ21_DT_survey}, and its formal definition was given in the white paper in 2017 \cite{grieves2017digital}. 
In recent years, DT has emerged as a significant strategic technology trend, with extensive investigations across various industries such as manufacturing operations, healthcare services, and urban planning \cite{AliPBHSS21, WickramasingheJ22}.
It is anticipated that 6G and DT will establish a mutually beneficial partnership \cite{KhanHSHGH22, TrungQDuong_WC23}. 
DT for wireless networks can be leveraged to coordinate computation and communication resources, thereby serving users with diverse requirements. Additionally, advanced communication techniques can provide seamless and low-latency connections to support the need of DT signaling.

Recent studies have explored the potential of DT-enabled mobile edge computing (MEC) as a typical application in 6G owing to the innumerable advantages of DT, as demonstrated in \cite{DITEN, BinLi_DT_TVT22, ZhangLong_DT_TVT23}. 
MEC plays a crucial role as an infrastructure component in current 5G and forthcoming 6G networks. It enables devices with limited computing resources to offload their tasks to nearby servers, thereby reducing the delay in completing computation tasks and improving the resource utilization efficiency \cite{MaoYZHL17, MinghuiDai_TITS21}. 
MEC has been widely studied from various perspectives such as computation rate maximization \cite{SuzhiBi_TWC18, GuangjiChen_MEC_JSAC23}, energy consumption minimization \cite{MinghuiDai_TNSE23}, and latency minimization \cite{ZhaoHLS23, Delay_Minimization_TCOM20, WuNZQT18}.
Based on these prior works, MEC can be employed to enable numerous delay-sensitive applications, including but not limited to, face recognition and indoor security surveillance, as well as automatic driving \cite{MaoYZHL17}.
Moreover, multi-tier computing (MTC) has emerged as an innovative and efficient computing architecture that can facilitate the scheduling of intensive tasks to heterogeneous servers located in different tiers, thereby increasing computing efficiency \cite{KunlunWang_cellfree_JSAC23, URLLC_JSAC23}. Recent studies have focused on enabling flexible coordination of computation, storage, and communication resources in MTC \cite{KunlunWangYZNTJ_JSAC23}, leading to the proposal of sophisticated offloading strategies.
For instance, joint task offloading and caching were studied in \cite{KunlunWang_TCOM22} which facilitates the task offloading from UEs to the terminal fog server via a relay equipped with massive antennas. Besides, the non-orthogonal multiple access (NOMA) technique has been exploited in \cite{ZhaolinWang_NOMAMTC_JSAC23} to support both target sensing and multi-tier task offloading.

Digital twin edge network (DITEN) incorporates DT and MEC into an integrated system \cite{DITEN}.		
In this integrated system, MEC facilitates flexible computing for DT decision-making, while DT monitors the real-time system status and precisely allocates communication and computing resources for MEC.
Such an integration enables the monitoring and prediction of the computing state of MEC systems, e.g., the central processing unit (CPU) clock frequency.
DITEN has garnered significant recognition as a pivotal catalyst in the achievement of Industry 4.0 \cite{ZhangLong_DT_TVT23}, making it a crucial architecture in modern industrial applications.
The integration of MEC and DT enables the real-time and efficient monitoring of network status information, thereby facilitating informed decision-making within DITEN systems.
This integration ensures the optimal utilization of system resources and establishes it as a highly efficient and effective solution for industrial applications.
In recent studies such as \cite{LiuTWCZ22, DuyVDCD22}, DT has been employed to assist edge server selection and task offloading for UEs. 
Deep reinforcement learning (DRL) was harnessed to minimize the energy consumption of an aerial DITEN via jointly optimizing unmanned aerial vehicle trajectory, transmission power and computation resource allocation in \cite{BinLi_DT_TVT22}. Furthermore, The Lyapunov technique was combined to reduce the computation latency in a time-varying MEC environment  \cite{Latency_DRL_TVT20}.
In \cite{RuiDong_DT_TWC19}, a digital twin that duplicates the network environment was established for offline training, and then the user association in the physical world can be designed in a real-time manner.

As discussed above, it is widely recognized that the integration of MEC/MTC and DT in the forthcoming 6G network is poised to revolutionize the current computing system into a mutually beneficial one.
Recent studies have emphasized the benefits that DT can provide to edge networks, such as real-time monitoring, prediction capability, and virtual simulation with lower trial-and-error costs.
However, it is essential to note that to maintain precise modeling of the entire system and allocate available resources accurately, DT may require frequent interaction with the physical world.
6G is envisioned to enable intelligent hyperconnectivity for a vast number of devices in DT applications, the development of efficient signaling strategies remains an open question that has not yet been addressed.
Recently, we have witnessed the emergence of a task-oriented multiple-access (MA) strategy called over-the-air computation (AirComp), which has shown remarkable efficiency in data aggregation \cite{AirComp_survey, FL_IRS23}. 
AirComp leverages co-channel interference to compute desired functions. This approach enables all devices to access the same channel, where signals can be naturally summed over the air. 
By adopting suitable pre- and post-process functions, AirComp can realize one-shot function computation, such as weighted sum, Euclidean norm, and maximum value. This makes AirComp an exceptionally promising solution for enabling efficient system monitoring, and it has been validated to be a scalable strategy in \cite{YuanmingShi_TWC21_FL_IRS, MultiCluster_AirComp}.
However, AirComp's performance is inevitably limited by unfavorable propagation paths which may be blocked by some obstacles like buildings and hills \cite{FL_IRS23}. This can lead to significant energy decaying on signals, especially when operating on the high frequency domains, e.g., millimeter and terahertz waves.
As a remedy, intelligent reflecting surface (IRS) emerges as a cost-effective technique for dynamically altering wireless channels via tuning signal reflections over low-cost passive reflecting elements.
Recent works have shown the promise of IRSs in ameliorating the performance of MEC \cite{RIS_NOMA_MEC_TCOM23, GuangjiChen_WPMEC_TWC23, GuangjiChen_MEC_JSAC23} and AirComp \cite{FangJSZCL21, MultiCluster_AirComp}. 
Specifically, a unified dynamic IRS beamforming framework was proposed in \cite{GuangjiChen_MEC_JSAC23} to enhance the computing rate via jointly optimizing the task offloading policy, time allocation, and the IRS phase shift configuration. 
Furthermore, the MEC system under amalgamated assistance of IRS and massive MIMO relay was studied in \cite{WangZWCY22}. Regarding the IRS-aided AirComp system, single-cluster and multi-cluster scenarios were respectively studied in \cite{FangJSZCL21, MultiCluster_AirComp}, which demonstrated that the IRS can significantly improve the computation accuracy of AirComp.

Motivated by the above discussions, we study in this paper a multi-tier hybrid computing system assisted by an IRS to achieve mutual benefits of DT and MEC by jointly considering the task offloading assisted by DT as well as the system monitoring for DT.
We embark upon the joint optimization of the offloading strategies of MEC UEs that are facilitated by DT, while concurrently ensuring the meticulous monitoring within the DT system through the ingenious utilization of the AirComp technique.
Although previous studies have investigated the separate design of DT-enabled MEC and AirComp, the network-wide convergence of these technologies is still a nascent field that requires further exploration. 
It is within this context that the primary objective of our work emerges, as we endeavor to construct a seamlessly integrated system that amalgamates DT and MEC while discovering the potential role of the IRS. 
One of the significant challenges encountered in the development of such a system lies in the inherent complexity arising from the superposition of signals emanating from diverse types of user equipment (UE) that are multiplexed on the same channel. 
This intricate complexity engenders critical issues unique to integrated networks, distinct from those present in conventional networks consisting solely of AirComp or MEC UE.
The main contributions of this paper can be summarized as follows.
\begin{itemize}
	\item For the first time, we propose a novel DITEN framework by considering the mutual effect of DT and MEC functionalities. Specifically, the tasks from MEC UE can be either locally processed, or offloaded and then computed in the servers located in higher tiers. Besides, the server in the middle tier is also served as a fusion center to aggregate data from distributed UEs via the AirComp technique. We formulate the corresponding hybrid computation rate maximization problem under the communication and computation resources constraints as well as the communication-computation causality.
	\item A major challenge in this problem is how to distinguish the homogeneous signals received from different types of UEs.
	To address the communication and computation constraints for task computing, MEC UE offloading, and AirComp functionality, it is imperative to devise an inter-layer association strategy. Additionally, a precise configuration of the IRS is required to facilitate signal transmission from heterogeneous UEs to the ES, while mitigating the interference among users.
	The optimization problem formulated in this paper is non-convex and NP-hard due to the highly coupled optimization variables and the combinatorial nature of the multiple devices' task offloading selection.
	We first analyze its inherent characteristics to simplify the original problem, then induce various auxiliary variables to equivalently transform the original optimization problem into a more tractable one. Further, we utilize the alternative optimization (AO) and successive convex approximation (SCA) methods to construct a series of convex optimization problems to obtain the local optimal solution for the original problem. 
	\item Simulation results demonstrate the effectiveness of the proposed algorithm for optimizing communication, computation, and time resources allocation. It is observed that the employment of IRS is beneficial for signal transmission in different tiers, and thereby can significantly enhance the performance of DITEN. Furthermore, numerical results show that the integration of DT and MEC can support precise task offloading strategy since DT can offer real-time system status. Throughout this paper, we emphasize the importance of DT in the integration of DT and MEC and demonstrate its necessity for achieving optimal performance in DITEN systems.
\end{itemize}

The remainder of this paper is organized as follows. 
In Section II, we describe the proposed IRS-aided multi-tier hybrid computing framework for the DITEN that includes DT, communication, and computation models. Then, we formulate the corresponding sum computation rate maximization problem under resources and causalities constraints.
In Section III, we analyze the original problem and then propose an efficient algorithm based on AO and SCA techniques.
Numerical results are presented in Section IV to validate the proposed algorithm, and the paper is concluded in Section V finally.

\emph{Notation:} Scalars are denoted by italic letters, and vectors and matrices are denoted by bold-face lower-case and uppercase letters, respectively. $\mathbb{R}^{m\times n}$ and $\mathbb{C}^{m\times n}$ denote the space of $m\times n$ real-valued and complex-valued matrices, respectively. For a complex-valued vector $\mathbf{x}$, ${\left\| \mathbf{x} \right\|}$ and ${\rm arg}(\mathbf{x})$ denote the corresponding Euclidean norm and its phase, respectively. Besides, ${\rm diag}(\mathbf{x}) $ denotes the diagonal matrix in which the main diagonal elements are extracted from the vector $\mathbf{x}$. For a square matrix $\mathbf{S}$, $\mathrm{tr}(\mathbf{S})$ denotes its trace. ${\mathbf{S}} \succeq {\mathbf{0}}$ means that the square matrix $\mathbf{S}$ is positive semi-definite. For any general matrix $\mathbf{A}$, $\mathbf{A}^{H}$, $\mathrm{rank}(\mathbf{A})$, and $\mathbf{A}_{i,j}$ denote its conjugate transpose, rank, and $(i, j)$th entry, respectively. $\mathbf{I}_M$ represents the identity matrix of size $M\times M$. $\jmath$ denotes the imaginary unit, i.e., $\jmath^2= -1$. $\mathbb{E}[\cdot] $ denotes the statistical expectation. 
${\cal O}\left(  \cdot  \right)$ is the big-O computational complexity notation.

\section{System Model and Problem Formulation}

\begin{figure}[t]
	\centering
	\includegraphics[width=1\columnwidth]{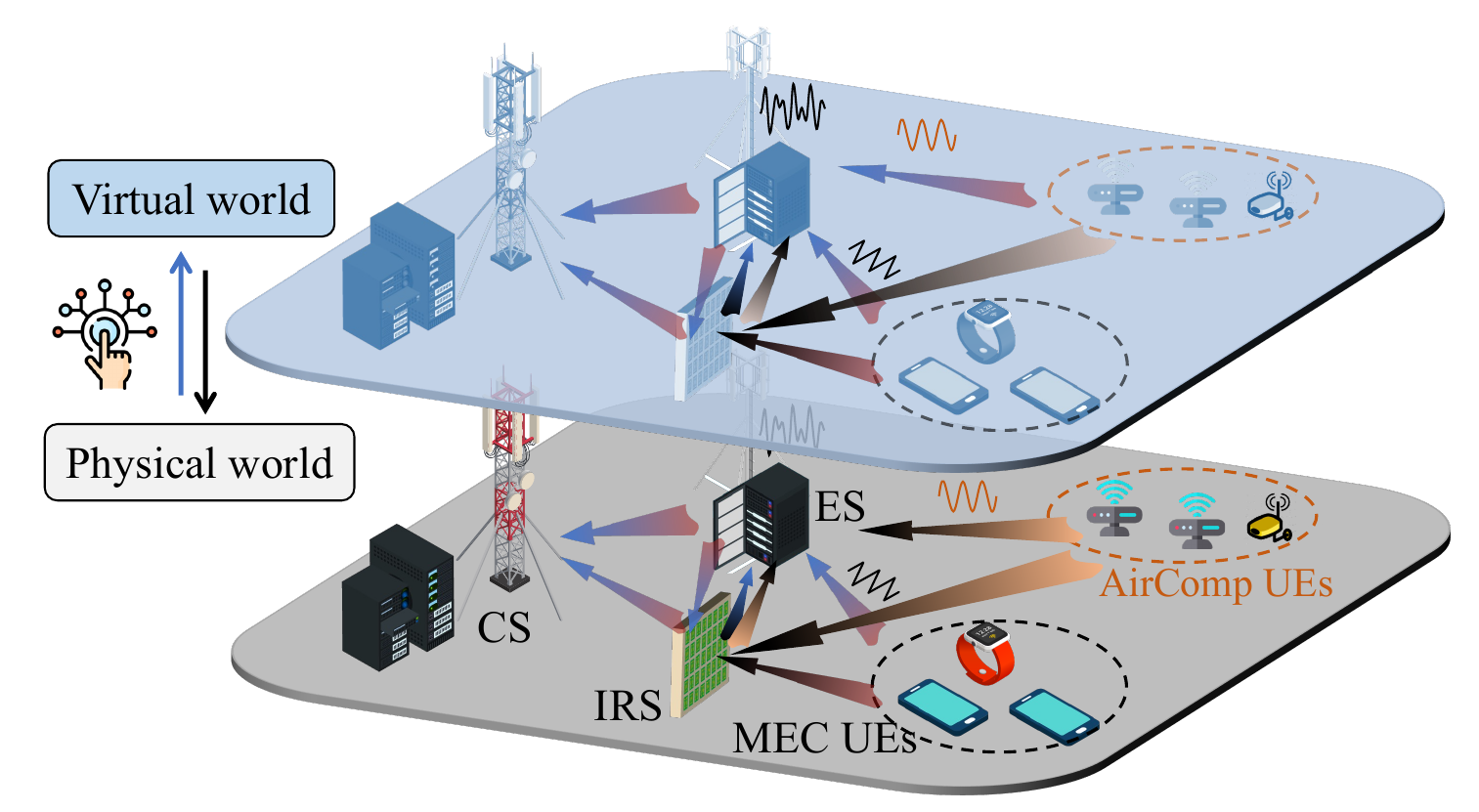}
	\caption{ IRS-aided multi-tier hybrid computing model for DITEN.}
	\label{fig_model}
\end{figure}

\begin{figure}[t]
	\centering
	\includegraphics[width=0.8\columnwidth]{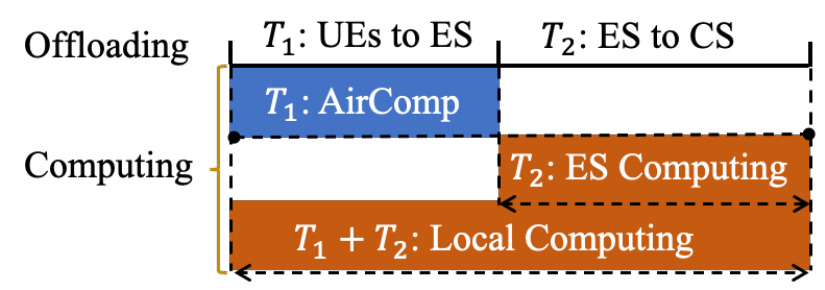}
	\caption{The hybrid computing protocol.}
	\label{protocol}
\end{figure}

As shown in Fig. 1, we consider an IRS-aided multi-tier hybrid computing system for DITEN, including both AirComp and MEC capabilities.
Specifically, the network contains multiple single-antenna hybrid UEs in the first tier, a base station (BS) equipped with $M_1$ antennas that serves as an edge server (ES) in the second tier, as well as a cloud server (CS) with $M_2$ antennas in the third tier. Besides, an IRS equipped with $N$ passive reflecting elements is deployed near the ES to support the signal transmissions among different tiers. Note that the hybrid UEs are composed of two types, i.e., the AirComp UE $k \in \mathcal{K}_a$ and the MEC UE $k \in \mathcal{K}_o$. 

AirComp UE seeks to monitor the status of the DT system, while MEC UE offloads heavy computing tasks to the higher tiers. All the UEs simultaneously upload their data to the ES over the same wireless channel with the assistance of an IRS to perform system monitoring and task offloading. 
We assume that all the devices are perfectly synchronized according to the synchronization techniques mentioned in \cite{AirComp_survey}.
In the ES, the successive interference cancellation (SIC) method is employed to sequentially decode the overlapped signals from different UEs. 
Without loss of generality, we adopt the partial data offloading strategy in the considered system and it is assumed there are no direct communication links between the UEs and the CS. The hybrid computing protocol is depicted in Fig. \ref{protocol}, which contains two periods $T_1$ and $T_2$. In the first hop during time period $T_1$, the MEC UE can either process tasks locally or upload them to an ES for remote execution. Meanwhile, the ES monitors the DITEN system via the AirComp technique during $T_1$. In the second period $T_2$, ES can further offload tasks to the CS for faster computing after receiving the tasks offloaded from UEs in $T_1$. Since the CS is usually equipped with sufficient computing capability, the corresponding computation time in the CS can be deemed negligible.
All the signal transmissions are facilitated through the IRS that by strategically reflecting signals to mitigate interference and improve signal strength.
By balancing local computing and task offloading strategy under the assistance of DT, the overall computing capabilities of the DITEN can be fully exploited.
In the subsequent, we introduce the considered DITEN system from DT, communication, and computation model respectively for a more comprehensive understanding.

\subsection{DT Model}
The DT layer is able to virtually replicate the physical entities including UEs and servers, and exhibit the information such as historical data and hardware impairment for centralized coordination.
For the $k$-th UE, its DT counterpart can be expressed as $\mathrm{DT}_k^{\mathrm{lo}} = (f_k^{\mathrm{lo}}, \hat{f}_k^{\mathrm{lo}})$, where $f_k^{\mathrm{lo}}, \hat{f}_k^{\mathrm{lo}}$ are the estimated CPU frequency assigned to the local task in the physical UE and the deviation between the value in the real UE and its DT, respectively. The actual CPU frequency can be expressed as $f_{\mathrm{true}} = f_{k}^{\mathrm{lo}} - \hat{f}_{k}^{\mathrm{lo}}$. Similarly, the DT counterparts for ES and CS can be expressed as $\mathrm{DT}^{\mathrm{es}} = (f^{\mathrm{es}}, \hat{f}^{\mathrm{es}})$ and $\mathrm{DT}^{\mathrm{cs}} = (f^{\mathrm{cs}}, \hat{f}^{\mathrm{cs}})$, respectively. 


\subsection{Communication Model}

We assume that all nodes in this system operate on the same frequency band, and the channels exhibit quasi-static flat-fading characteristics, i.e., the channel coefficients remain constant within a given coherence time. 
The IRS can be configured sequentially in $T_1$ and $T_2$ to assist the signal transmission among different tiers. Let $\mathbf{\Theta}_1$ and $\mathbf{\Theta}_2$ denote the passive beamforming pattern of IRS in $T_1$ and $T_2$, respectively. 
The baseband channels from the IRS to ES, from the MEC UE $k$ to IRS, and from the MEC UE $k$ to ES are denoted by $\mathbf{G}^{H}_e \in \mathbb{C}^{M_1 \times N}$, $\mathbf{f}_{o,k} \in \mathbb{C}^{N \times 1}$ and $\mathbf{h}_{o,k} \in \mathbb{C}^{M_1 \times 1}$, respectively.
Hence, the composite MEC channels are given by $\tilde{\mathbf{h}}_{o,k} = \mathbf{h}_{o,k} + \mathbf{G}^{H}_e \mathbf{\Theta}_1 \mathbf{f}_{o,k} = \mathbf{h}_{o,k} + \mathbf{G}^{H}_e \mathrm{diag}(\mathbf{f}_{o,k}) \mathbf{v}_1,\; \forall k \in \mathcal{K}_o$, where $\mathbf{\Theta}_1 = \mathrm{diag}(\mathbf{v}_1)$, $\mathbf{v}_1= [v_{1,1}, \ldots, v_{1,N}]^T$. Each entry $v_{1,n} = e^{\jmath \theta_n}, 1\leq n \leq N$ in $\mathbf{v}_1$ denotes the phase tuning of each element of IRS. 
Similarly, the composite AirComp channels are given by $\tilde{\mathbf{h}}_{a,k} = \mathbf{h}_{a,k} + \mathbf{G}^{H}_e \mathbf{\Theta}_1 \mathbf{f}_{a,k} = \mathbf{h}_{a,k} + \mathbf{G}^{H}_e \mathrm{diag}(\mathbf{f}_{a,k}) \mathbf{v}_1,\; \forall k \in \mathcal{K}_a$. 
 Besides, the composite ES-CS channel is given by $\tilde{\mathbf{H}}_{c} = \mathbf{H}_{c} + \mathbf{G}^{H}_c \mathbf{\Theta}_2 \mathbf{F}_{c}$, where $\mathbf{G}^{H}_c$, $\mathbf{F}_{c}$, and $\mathbf{H}_{c}$ denote the baseband channels from the IRS to CS, from the ES to IRS, from the ES to CS,  respectively, and $\mathbf{\Theta}_2 = \mathrm{diag}(\mathbf{v}_2)$ with $\mathbf{v}_2= [v_{2,1}, \ldots, v_{2, N}]^T$ denotes the phase shift of IRS in the second stage. We assume that the channel state information is perfectly obtained via the existing channel estimation techniques \cite{WQQ_IRS_Survey}.
 
The communication process in DITEN is composed of two hops. One is the hybrid UEs transmit system monitoring data and offload tasks to the ES in the first stage $T_1$, and the other is the task offloading from ES to the CS in the second stage $T_2$. Further details are presented below.
\subsubsection{First Stage $T_1$} 
The received signal at the ES during $T_1$ is comprised of the monitoring signal from AirComp UEs $k \in \mathcal{K}_a$ and the task offloading signal from MEC UEs $k \in \mathcal{K}_o$, that is given by
\begin{align}
	\mathbf{y}_{e} = \sum_{k \in \mathcal{K}_a}  \tilde{\mathbf{h}}_{a,k} b_{k} s_{a,k}+ \sum_{k\in \mathcal{K}_o} \tilde{\mathbf{h}}_{o,k} \sqrt{p_k} s_{o,k} + \mathbf{n}_e,
\end{align}
where $b_{k} \in \mathbb{C}$ is the transmit scalar of AirComp users, $\sqrt{p_k}$ is the transmit power of task offloading users, and $\mathbf{n}_e\in \mathbb{C}^{M_1\times1}$ denotes the additional white Gaussian noise at ES that is distributed as $\mathcal{CN}(0, \sigma_e^2 \mathbf{I}_{M_1})$. Besides, $s_{a,k}$ and $s_{o,k}$ denote the transmit symbol of AirComp UEs and MEC UEs, respectively. Without loss of generality, we assume that each symbol has been normalized into a mutually independent symbol with zero mean and unit power.

After receiving the transmitted signals from various UEs, the ES performs SIC to decode the signals from different UEs. Without loss of generality, we assume that the ES first decodes the superimposed signal from the AirComp UEs $k \in \mathcal{K}_a$ and then subtracts them to further decode the data from task offloading users. 
This is due to that the signals from all the AirComp UEs need to be decoded simultaneously to realize one-shot function computation, and the combined strength of the superimposed signal from all the AirComp UEs is likely to be larger than each signal from the task offloading UE. Hence, we assume that the ES firstly decodes the AirComp signal and then sequentially decodes the individual signal from task offloading UEs.
By regarding the task offloading signals as noise, the ES first decodes the superimposed signal from the AirComp UEs $k \in \mathcal{K}_a$.
Given the decoder $\mathbf{a} \in \mathbb{C}^{M_1\times1}$, the recovered signal at ES can be expressed as
\begin{align}
	\mathbf{a}^H \mathbf{y}_{e} =& \sum_{k \in \mathcal{K}_a} s_{a,k} + \sum_{k \in \mathcal{K}_a} ( 	\mathbf{a}^H \tilde{\mathbf{h}}_{a,k} b_{k} - 1 ) s_{a,k} \nonumber \\ 
	&+ \sum_{\forall k \in \mathcal{K}_o} \mathbf{a}^H \tilde{\mathbf{h}}_{o,k} \sqrt{p_k} s_{o,k} + \mathbf{a}^H  \mathbf{n}_e.
\end{align}
Hence, the corresponding MSE of decoding the AirComp signal is given by
\begin{align}
	\!\mathrm{MSE}_{a} \! = \! \! \! \sum_{k \in \mathcal{K}_a} \! \! |	\mathbf{a}^H \tilde{\mathbf{h}}_{a,k} b_{k} \! - \! 1|^2 \! + \! \! \sum_{k \in \mathcal{K}_o} \! p_k |\mathbf{a}^H \tilde{\mathbf{h}}_{o,k} |^2 \! \! + \! \| \mathbf{a}\|^2 \sigma_e^2.\!\!
\end{align}

Then, the ES subtracts the recovered AirComp signal from $\mathbf{y}_e$ and decodes the signals from the task offloading UEs as the traditional uplink NOMA. With a given decoding order, the ES first decodes the signals of device $i, \forall i <k $ then subtracts them from the superimposed signal for further decoding \cite{MinFu_NOMA}. 
The corresponding SINR for decoding the signal from task offloading UE $k$ is given by 
\begin{align}
	\gamma^e_{k} = \frac{p_k\|\tilde{\mathbf{h}}_{o,k}\|^2 }{\sum_{i>k} p_i \| \tilde{\mathbf{h}}_{o,i} \|^2 + \sigma^2_{e} } ,\ \forall k \in \mathcal{K}_o,
\end{align}
and the task offloading rate (in bits/s/Hz) of UE $k$ is given by
\begin{align}
	r^{e}_k =\log_2 \big(1+ \gamma^e_{k}  \big).
\end{align}
Thereby, the sum offloading rate is given by 
\begin{align}
	\sum_{k\in \mathcal{K}_o} r^{e}_k = \log_2 \big(1+ \frac{ \sum_{k\in \mathcal{K}_o} p_k\|\tilde{\mathbf{h}}_{o,k}\|^2 }{  \sigma^2_{e} }  \big).
\end{align}

\subsubsection{Second Stage $T_2$}

The ES can further offload the task to CS during $T_2$ to acquire a faster computing rate. Denote $ \mathbf{H}_{c} \in \mathbb{C}^{M_2 \times M_1} $, $\mathbf{G}_c \in \mathbb{C}^{M_2 \times N} $, $\mathbf{F}_{c} \in \mathbb{C}^{N \times M_1}$ as the baseband channels from the ES to CS, from the IRS to CS, and from the ES to IRS, respectively. Thereby, the composite channel from ES to CS is given by $\tilde{\mathbf{H}}_{c} = \mathbf{H}_{c} + \mathbf{G}_c \mathbf{\Theta}_2 \mathbf{F}_{c}$.
The received signal at CS is presented as
\begin{align}
	\mathbf{y}_c = \tilde{\mathbf{H}}_{c} \mathbf{s}_e + \mathbf{n}_c,
\end{align}
where $\mathbf{s}_e \in \mathbb{C}^{M_1 \times 1}$ is the transmit signal vector at ES, and $\mathbf{n}_c\in \mathbb{C}^{M_2\times1}$ denotes the additional white Gaussian noise at ES that is distributed as $\mathcal{CN}(0, \sigma_c^2 \mathbf{I}_{M_2})$. 
Let $\mathbf{W} = \mathbb{E}[\mathbf{s}_e \mathbf{s}_e ^H]  \succeq \mathbf{0}  \in \mathbb{C}^{M_1 \times M_1}$ denote the transmit signal covariance matrix.
The task offloading rate (in bits/s/Hz) from ES to the CS is thus given by \cite{ShuowenZhang_MIMO_IRS}
\begin{align}
	r^{c} = \log_2 \mathrm{det} (\mathbf{I}_{M_2} + \frac{1}{\sigma^2_c} \tilde{\mathbf{H}}_{c} \mathbf{W}  \tilde{\mathbf{H}}_{c}^H).
\end{align}

\subsection{Computation Model}
The computation model in the considered multi-tier hybrid computing system is divided into two parts, one is the computation for system monitoring via AirComp in the first stage $T_1$, and the other is the task computing in the MEC system during the whole period $T$.

\subsubsection{AirComp Rate}
The computation rate of AirComp measured in bits was proposed in \cite{Nazer_TIT11}, and has been studied in \cite{MultiCluster_AirComp, AirComp_survey}. The explicit form  is given by
\begin{align}
	R^{\mathrm{a}} = \mathrm{log}_2^+ \big(1/\mathrm{MSE}_{a} \big),
\end{align}
where $\mathrm{log}_2^+(\cdot) = \max\{\mathrm{log}_2(\cdot), 0\}$. Since AirComp is utilized to monitor the DITEN system, generally $R^{\mathrm{a}}>0$ is needed.

\subsubsection{MEC Rate}
The computation rate in the MEC system can be further divided into three parts corresponding to different tiers.
Let $\rho$ denote the number of central processor unit (CPU) cycles required for completing $1$-bit data, the local computation rate, i.e., the number of bits processed per second, of user $k$ is given by
\begin{align}
	R^{\mathrm{lo}}_k = \frac{f^{\mathrm{lo}}_k - \hat{f}_k^{\mathrm{lo}}}{\rho}.
\end{align}
Similarly, the local computation rate at the ES is given by
\begin{align}
	R^{\mathrm{es}} = \frac{f^{\mathrm{es}} - \hat{f}^{\mathrm{es}}}{\rho}.
\end{align}
We assume that the CS has sufficient computing capability, thus the corresponding computation rate is deemed as infinite and out of consideration.

\subsection{Power Consumption Model}
As described above, there exist heterogeneous nodes and hierarchical computing networks in the considered intelligent computing system. We consider the nodes to be power-limited, and their power need to be precisely optimized to maximize the hybrid computation rate. 
In the first stage $T_1$, both the AirComp users and the task offloading users simultaneously transmit their data to the ES with the assistance of an IRS.  We have the following power constraints
\begin{align}
	& |b_{k}|^2  \leq P_{a},\forall k \in \mathcal{K}_{a} , \label{} \\
		& p_k + \kappa(f^{\mathrm{lo}}_{k,1} \! - \! \hat{f}_{k}^{\mathrm{lo}})^3 \leq P_{o}, \forall k \in \mathcal{K}_{o},
\end{align}
where $f^{\mathrm{lo}}_{k,1}$ denotes the assigned CPU frequency of UE $k$, and $\kappa$ is the computational energy efficiency of CPU chip.
In the second stage $T_2$, the MEC UEs and the ES can perform local computing while the ES can further offload the tasks to CS. The ES needs to assign its power to balance local computing and task offloading. The corresponding power constraints in $T_2$ are given by 
\begin{align}
	& \kappa(f^{\mathrm{lo}}_{k,2}  -  \hat{f}_k^{\mathrm{lo}})^3 \leq  P_{o}, \forall k \in \mathcal{K}_{o}, \label{}\\
	& \mathrm{tr}(\mathbf{W}) + \kappa(f^{\mathrm{es}} - \hat{f}^{\mathrm{es}})^3 \leq   P_{\mathrm{es}},
\end{align}
where $\{f^{\mathrm{lo}}_{k,2}\}, f^{\mathrm{es}}$ denote the assigned CPU frequency of MEC UEs and the ES, $\mathrm{tr}(\mathbf{W}) $ represents the transmit power consumption of the ES to CS.

The overall computation rate over $T$ is composed of the AirComp rate within $T_1$ and the MEC rate in $T$. 
Since the CS is usually equipped with sufficient computing capability, the corresponding computing time can be deemed negligible.
Therefore, the MEC rate can be presented as the summation of the local computation rate at UEs and the task offloading rate from UEs to the ES, with the explicit formation given by 
\begin{align}
	 	R_{\mathrm{MEC}}  = \sum_{k\in \mathcal{K}_o} \big( T_1 B r^{\mathrm{e}}_k  +  T_1R^{\mathrm{lo}}_{k,1}  +  T_2R^{\mathrm{lo}}_{k,2}).
\end{align}
The weighted sum computing rate contains AirComp rate and MEC rate is given by
\begin{align} \label{hybrid_sumrate}
	R_{\mathrm{total}} = w_a T_1 R^a  \! + \! w_o \! \! \sum_{k\in \mathcal{K}_o} \! \! \!\big( T_1 B r^{\mathrm{e}}_k  \! + \! T_1R^{\mathrm{lo}}_{k,1} \!+\! T_2R^{\mathrm{lo}}_{k,2}),
\end{align}
where $w_a$ and $w_o$ denote the weights of AirComp rate and MEC rate, respectively. 

Furthermore, the MEC computation rate and task offloading rate must adhere to causality constraints due to their interdependent relationship. Specifically, the sum task offloading bits from MEC UEs to the ES  cannot exceed the summation of computation rate at the ES and  the offloading rate from the ES to the CS, the corresponding constraint is given by
\begin{align}
	\sum_{k\in \mathcal{K}_o} T_1 B r^{\mathrm{e}}_k \leq T_2 B r^{\mathrm{c}} + T_2 R^{\mathrm{es}}.
\end{align}

So far, we have presented the sophisticated multi-tier hybrid computing model for DITEN. 
We aim to optimize the time, communication, and computation resources to maximize the hybrid computation rate while satisfying the resource constraints as well as the communication-computation causality. MEC UE and the ES need to allocate their power to balance the local computing and task offloading capability as well as to reduce the interference to AirComp UEs. 
The corresponding problem is formulated as
\begin{subequations} \label{P1}
	\begin{align}
		\!\! \!\! \!\! \mathop  {\max}\limits_{\scriptstyle \{p_k\}, \mathbf{W}, \{b_{k}\}, \{T_i\}
			\atop \scriptstyle 
			\mathbf{a}, \{\mathbf{\Theta}_i\}, \{f_{k,1}^{\mathrm{lo}},f_{k,2}^{\mathrm{lo}},f^{\mathrm{es}}\} } & R_{\mathrm{total}}  \label{P1_obj} \\
		{\rm{s.t.}}\ \ \ \ \ \ \ \ \
		&\!R^{\mathrm{a}} > 0, \label{P1_AirCompRate} \\
		&\!T_1 \!+ T_2 \leq T, \label{P1_c_time} \\
		&\!\sum_{k\in \mathcal{K}_o} T_1 B r^{\mathrm{e}}_k \leq T_2 B r^{\mathrm{c}} + T_2 R^{\mathrm{es}}, \label{P1_c_bits} \\
		&\!|b_{k}|^2  \leq P_{a},\; \forall k \in \mathcal{K}_{a} , \label{P1_c_AirComp} \\
		&\! p_k + \kappa(f^{\mathrm{lo}}_{k,1} \! - \! \hat{f}_{k,1}^{\mathrm{lo}})^3 \leq P_{o}, \; \forall k \in \mathcal{K}_{o}, \label{P1_c_UE1}\\
		&\!\kappa(f^{\mathrm{lo}}_{k,2} \! - \! \hat{f}_{k,2}^{\mathrm{lo}})^3 \! \leq \! P_{o},\; \forall k \in \mathcal{K}_{o}, \label{P1_c_UE2}\\
		&\!\mathrm{Tr}(\mathbf{W}) +  \kappa(f^{\mathrm{es}} \! -\! \hat{f}^{\mathrm{es}})^3   \! \leq  \!  P_{\mathrm{es}},  \! \label{P1_c_ES}\\
		&\!f_{k,1}^{\mathrm{lo}} - \hat{f}_k^{\mathrm{lo}} \leq F_{\mathrm{lo},k}^{\max},\; \forall k \in \mathcal{K}_{o}, \label{P1_c_UE_DT} \\
		&\!f_{k,2}^{\mathrm{lo}} - \hat{f}_k^{\mathrm{lo}} \leq F_{\mathrm{lo},k}^{\max},\; \forall k \in \mathcal{K}_{o}, \label{P1_c_UE_DT1} \\
		&\!f^{\mathrm{es}} - \hat{f}^{\mathrm{es}} \leq F_{\mathrm{es}}^{\max},  \label{P1_c_ES_DT} \\
		&\!\theta_{i,n} \in (0,2\pi], \ \forall i,n. \label{P1_c_IRS} 
	\end{align}
\end{subequations} 
The weights in the objective function \eqref{P1_obj} can be adjusted to reflect different priorities and enforce fairness principles among devices.
However, since these weights do not impact the algorithm design, we assume equal weighting in this paper for the sake of simplicity and without loss of generality, i.e., $w_a=w_o=0.5$.
In problem \eqref{P1}, 
\eqref{P1_AirCompRate} is to ensure that AirComp is always available to support the DT system, \eqref{P1_c_time} is the time constraint of total task computing and offloading, \eqref{P1_c_bits} is the communication-computation causality constraint to guarantee that the task offloaded from UEs can be successfully computed, i.e., the computing bits at ES and offloading bits from ES to CS are large than the task offloading rate bits from UEs. Besides, \eqref{P1_c_AirComp}, \eqref{P1_c_UE1}, \eqref{P1_c_UE2}, and \eqref{P1_c_ES} are the power constraints of AirComp UEs $k\in \mathcal{K}_a$, task offloading UEs $k \in \mathcal{K}_o$ and the ES. \eqref{P1_c_UE_DT}, \eqref{P1_c_ES_DT} are the computing resource constraints at UEs and ES, and \eqref{P1_c_IRS} denotes the phase shift constraint of IRS. Problem \eqref{P1} is non-convex and NP-hard since the optimization variables are highly coupled in the objective function and the constraints. Besides, it is noted that the problem may be infeasible due to the AirComp rate requirement \eqref{P1_AirCompRate}. We first assume that problem \eqref{P1} is feasible and investigate its inherent properties, then propose an efficient algorithm based on AO and SCA techniques in the following section.
Furthermore, we study the feasibility of this problem, and then analyze its convergence behavior and computational complexity.

\section{Proposed Algorithm}

Given the phase shift configuration of IRS, the channel coefficients among the different tiers in DITEN are determined. However, problem \eqref{P1} remains intractable due to the multiplicative terms present in both the objective function \eqref{P1_obj} and the constraint \eqref{P1_c_bits}. To address this difficulty, we need to decouple the variables and then reformulate these terms into a more tractable form.
In this section, we first divide all the variables into three blocks, i.e., 1) transceiver design and computation resource allocation: $ \{p_k\}, \mathbf{W}, \{b_{k}\}, \mathbf{a}, \{f_{k,1}^{\mathrm{lo}},f_{k,2}^{\mathrm{lo}},f^{\mathrm{es}}\}$; 2) phase shift configuration of IRS in the two tiers: $\mathbf{v}_1, \mathbf{v}_2$; 3) time resource allocation to different tiers: $T_1, T_2$. 
We then introduce auxiliary variables to reformulate the non-convex subproblem for each block in an equivalent manner to handle the coupling among the optimization variables. 
Finally, we apply the SCA method to construct a series of convex optimization problems that are capable of obtaining local-optimal solutions for the original problem of each block.


\subsection{Transceiver Design and Computation Resource Allocation} \label{sub1}

In this part, we precisely design the transceiver in the considered DITEN system. We first simplify the problem by analyzing its inherent properties.
Focusing on the transceiver design in the first period $T_1$, we can obtain the following remark.
\begin{remark}
	Since the AirComp signal from $k\in \mathcal{K}_a $ is decoded at first, its signal strength is irrelevant to the task offloading signals transmitted from MEC UEs $k\in \mathcal{K}_o $. Therefore, the design procedure of $\{b_k\}$ can be separated from joint design with task offloading users.
\end{remark}

Motivated by Remark 1, we adopt the widely used uniform-forcing transceiver design of separated AirComp system \cite{MultiCluster_AirComp}. Let $\mathbf{a} = \frac{1}{\sqrt{\eta}}\mathbf{m}$, we have
\begin{align}
	&b_{k}  =  \sqrt{\eta}\frac{(\mathbf{m}^H \tilde{\mathbf{h}}_{a,k})^{H}}{|\mathbf{m}^{H}\tilde{\mathbf{h}}_{a,k}|^2}, \forall k \in \mathcal{K}_a, \\
	&\eta  =  P_{a} \min_k |\mathbf{m}^{H} \tilde{\mathbf{h}}_{a,k}^{H}|^2,  \forall k \in \mathcal{K}_a.
\end{align}
In that case, the corresponding MSE for decoding the AirComp signal is given by
\begin{align}
	\mathrm{MSE}_{a} = \frac{\sum_{k \in \mathcal{K}_o} p_k |\mathbf{m}^H \tilde{\mathbf{h}}_{o,k}|^2 + \|\mathbf{m}\|^2\sigma_e^2}{ P_{a}  \min_k |\mathbf{m}^{H} \tilde{\mathbf{h}}_{a,k}|^2}.
\end{align}
With constraint \eqref{P1_AirCompRate}, the computation rate of AirComp can be expressed as 
\begin{align}
	&R^{\mathrm{a}} = \mathrm{log}_2 \bigg(\frac{P_{a} \min_k |\mathbf{m}^{H} \tilde{\mathbf{h}}_{a,k}|^2}{\sum_{k \in \mathcal{K}_o} p_k |\mathbf{m}^H \tilde{\mathbf{h}}_{o,k} |^2 + \|\mathbf{m}\|^2\sigma_e^2} \bigg).
\end{align}
Hence, the corresponding problem with respect to the transceiver design and computation resource allocation is formulated as
\begin{subequations} \label{P2}
	\begin{align}
		\!\! \!\! \!\! \mathop  {\max}\limits_{\scriptstyle \mathbf{m}, \{p_k\}, \mathbf{W}, 
			\atop \scriptstyle 
		 \{f_{k,1}^{\mathrm{lo}},f_{k,2}^{\mathrm{lo}},f^{\mathrm{es}}\} } & R_{\mathrm{total}}  \label{} \\
		{\rm{s.t.}}\ \ \ \ \ \ \;
		& R^{\mathrm{a}} > 0, \\
		&\eqref{P1_c_bits},\eqref{P1_c_UE1},\eqref{P1_c_UE2},\eqref{P1_c_ES},\eqref{P1_c_UE_DT},\eqref{P1_c_UE_DT1},\eqref{P1_c_ES_DT}.
	\end{align}
\end{subequations}
Note that the objective function in \eqref{P2} is still very complex with highly coupled variables $\mathbf{m}$ and $\{p_k\}$.

By introducing three slack variables such that
\begin{align}
	&\frac{1}{S_a} =  \min_k |\mathbf{m}^{H} \tilde{\mathbf{h}}_{a,k}|^2,\; \forall k \in \mathcal{K}_a, \\
	&I_a = \sum_{k \in \mathcal{K}_o} p_k |\mathbf{m}^H \tilde{\mathbf{h}}_{o,k} |^2 + \|\mathbf{m}\|^2\sigma_e^2, \\
	&\frac{1}{S_o} = \frac{ \sum_{k\in \mathcal{K}_o} p_k\|\tilde{\mathbf{h}}_{o,k}\|^2 }{  \sigma^2_{e} }.
\end{align}
We can transform the objective function \eqref{P1_obj} to
\begin{align}
	R_{\mathrm{total}}  &=  w_a T_1 B \mathrm{log}_2 ( \frac{P_{a}}{S_a I_a} )  + w_o T_1 B \mathrm{log}_2 \big(1+ \frac{1}{S_o} \big) \nonumber \\
	& \quad + T_1R^{\mathrm{lo}}_{k,1} + T_2R^{\mathrm{lo}}_{k,2}.
\end{align}
In that case, problem \eqref{P1} can be converted to
\begin{subequations} \label{P21}
	\begin{align}
		\!\! \!\! \!\!\! \mathop  {\max}\limits_{\scriptstyle S_a, I_a, S_o, \mathbf{m}, \{p_k\},  
			\atop \scriptstyle \mathbf{W},
			  \{f_{k,1}^{\mathrm{lo}},f_{k,2}^{\mathrm{lo}},f^{\mathrm{es}}\} } \!\!\!\! & w_a T_1 B \mathrm{log}_2 ( \frac{P_{a}}{S_a I_a} )  + w_o T_1 B \mathrm{log}_2 \big(1+ \frac{1}{S_o} \big) \nonumber \\
	& \ + w_o(T_1R^{\mathrm{lo}}_{k,1} + T_2R^{\mathrm{lo}}_{k,2}) \label{P21_obj} \\
		{\rm{s.t.}}\ \ \;
		&\!\mathrm{log}_2 ( \frac{P_{a}}{S_a I_a} ) > 0, \label{P21_1} \\
		&\!\frac{1}{S_a} \leq  |\mathbf{m}^{H} \tilde{\mathbf{h}}_{a,k}|^2, \forall k \in \mathcal{K}_a, \label{P21_relax1} \\
		&\!I_a \geq \sum_{k \in \mathcal{K}_o} p_k|\mathbf{m}^H \tilde{\mathbf{h}}_{o,k} |^2 + \|\mathbf{m}\|^2 \sigma^2_{e}, \label{P21_relax2}\\
		&\!\frac{1}{S_o} \leq \frac{ \sum_{k\in \mathcal{K}_o} p_k\|\tilde{\mathbf{h}}_{o,k}\|^2 }{  \sigma^2_{e} }, \label{P21_relax3}\\
		&\!\eqref{P1_c_bits},\eqref{P1_c_UE1},\eqref{P1_c_UE2},\eqref{P1_c_ES},\eqref{P1_c_UE_DT},\eqref{P1_c_UE_DT1},\eqref{P1_c_ES_DT}.
	\end{align}
\end{subequations} 

\begin{remark}
	Note that problem \eqref{P21} is equivalent to problem \eqref{P2}. For the optimal solution of problem \eqref{P21}, constraints \eqref{P21_relax1}, \eqref{P21_relax2} and \eqref{P21_relax3} hold with equality. This is due to that we can always reduce the value of $S_a, S_o$ or increase the value of $I_a$ to increase the objective value until these constraints become active. Therefore, problem \eqref{P21} is equivalent to problem \eqref{P2}.
\end{remark}

However, problem \eqref{P21} is still a non-convex optimization problem due to the highly coupled variables in the objective function \eqref{P21_obj} and the constraints \eqref{P21_relax1}, \eqref{P21_relax2}, \eqref{P21_relax3}. In order to tackle this complex problem, we utilize the SCA technique to transform them into more tractable forms.

Note that both $\mathrm{log} ( \frac{P_{a}}{S_a I_a} )$ and $ \mathrm{log}_2 \big(1+ \frac{1}{S_o} \big)$ are jointly convex with respect to the variables $S_a$, $I_a$ and $S_o$. By applying the first-order Taylor expansion, the lower bounds at given local points $\{S_a^{(i)}, I_a^{(i)}, S_o^{(i)}\}$ in $i$-th iteration are given by
\begin{align}
	\mathrm{log}_2 ( \frac{P_{a}}{S_a I_a} ) &\geq  \mathrm{log}_2 ( \frac{P_{a}}{S_a^{(i)} I_a^{(i)}	} ) - \frac{\log_2 e}{S_a^{(i)}} (S_a-S_a^{(i)}) \nonumber \\
	&\quad - \frac{\log_2 e}{I_a^{(i)}}(I_a -I_a^{(i)}) \triangleq R^{\mathrm{low}}_a( S_a, I_a),
\end{align}
\begin{align}
	\mathrm{log}_2 \big(1+ \frac{1}{S_o} \big) &\geq \mathrm{log}_2 \big(1+ \frac{1}{S_o^{(i)}} \big) - \frac{(S_o-S_o^{(i)}) \log_2 e}{S_o^{(i)}(1+ S_o^{(i)})} \nonumber \\
	&\quad \triangleq R^{\mathrm{low}}_o (S_o).
\end{align} 
In that case, the objective function \eqref{P21_obj} is approximated as 
\begin{align}
	R_{\mathrm{total}}^{\mathrm{low}}  &\triangleq w_a T_1 B R^{\mathrm{low}}_a( S_a, I_a) + w_o T_1 B R^{\mathrm{low}}_o (S_o) \nonumber \\
	& \quad + T_1R^{\mathrm{lo}}_{k,1} + T_2R^{\mathrm{lo}}_{k,2}
\end{align}

It is observed that the remaining non-convexity of problem \eqref{P21} is the constraints \eqref{P21_relax1}, \eqref{P21_relax2}, \eqref{P21_relax3}.
To tackle this issue, we harness the SCA technique to convert the constraints into more tractable forms in the next.
Note that 
the right-hand-side (RHS) of \eqref{P21_relax1} is convex with respect to $\mathbf{m}$. 
Hence, their lower bound can be obtained via the SCA technique. For any given local point $\mathbf{m}^{(i)}$ in the $i$-th iteration, we have
\begin{align}
	|\mathbf{m}^{H} \tilde{\mathbf{h}}_{a,k}|^2 &\geq - (\mathbf{m}^{(i)})^H \tilde{\mathbf{H}}_{a,k} \mathbf{m}^{(i)} + 2\Re{(\mathbf{m}^{(i)})^H \tilde{\mathbf{H}}_{a,k} \mathbf{m}}, \nonumber \\
	&\triangleq f^{\mathrm{low}}_k (\mathbf{m},\mathbf{m}^{(i)}).
\end{align}

Note that the variables are highly coupled in constraints \eqref{P21_relax2}, \eqref{P21_relax3}, we induce a variable $t_k$ to  substitute $1/p_k$, i.e., $t_k = 1/p_k$ to convert them to more tractable forms which are given by
\begin{align}
	&I_a \geq \sum_{k \in \mathcal{K}_o} |\mathbf{m}^H \tilde{\mathbf{h}}_{o,k} |^2/t_k + \|\mathbf{m}\|^2\sigma_e^2, \label{P21_relax2_1} \\
	&\frac{1}{S_o} \leq  \sum_{k\in \mathcal{K}_o} \frac{ \|\tilde{\mathbf{h}}_{o,k}\|^2 }{ t_k \sigma^2_{e} }.\label{P21_relax3_1}
\end{align}
In that case, both the LHSs and RHSs of \eqref{P21_relax2_1} and \eqref{P21_relax3_1} are joint convex with respect to the optimization variables. By employing the SCA technique, we can obtain that
\begin{align} \label{P21_relax3_2}
\frac{1}{S_o} \leq \sum_{k\in \mathcal{K}_o} \bigg( \frac{ \|\tilde{\mathbf{h}}_{o,k}\|^2 }{ t_k^{(i)} \sigma^2_{e} } - \frac{\|\tilde{\mathbf{h}}_{o,k}\|^2(t_k - t_k^{(i)})}{(t_k^{(i)})^2 \sigma_e^2} \bigg),
\end{align}
where $\{t_k^{(i)}\}$ are the given local points in the $i$-th iteration.

Hence, problem \eqref{P21} can be tackled by solving a series of convex optimization problems, the corresponding problem in $i$-th iteration is given by

\begin{subequations} \label{P22}
	\begin{align}
		\!\! \!\! \!\! \mathop  {\max}\limits_{\scriptstyle S_a, I_a, S_o, \mathbf{m}, \{p_k\},  
			\atop \scriptstyle \mathbf{W},
			  \{f_{k,1}^{\mathrm{lo}},f_{k,2}^{\mathrm{lo}},f^{\mathrm{es}}\} } \! & R_{\mathrm{total}}^{\mathrm{low}} \\
		\!\!\!\!\!\!\!\!{\rm{s.t.}} \ \ \ \ \ \ \ \  
		&R^{\mathrm{low}}_a( S_a, I_a) > 0, \label{P22_Ra} \\
		&1/S_a \leq f^{\mathrm{low}}_k \! (\mathbf{m},\mathbf{m}^{(i)}),\forall k \in \mathcal{K}_a, \label{P22_Ra1}\\
		&1/t_k + \kappa(f^{\mathrm{lo}}_{k,1} \! - \! \hat{f}_{k,1}^{\mathrm{lo}})^3 \leq P_{o}, \forall k, \label{P22_Ra2}\\
		&\eqref{P1_c_bits},\eqref{P1_c_UE2},\eqref{P1_c_ES},\eqref{P1_c_UE_DT},\eqref{P1_c_UE_DT1},\eqref{P1_c_ES_DT}, \\
		&\eqref{P21_relax2_1}, \eqref{P21_relax3_2}.
	\end{align}
\end{subequations}
Problem \eqref{P22} is shown as a convex optimization problem that can be efficiently solved using standard convex programming solvers such as CVX. By properly inducing auxiliary variables to reformulate the non-convex objective function and constraints, and then approximating them via the SCA approach, problem \eqref{P21} can be ultimately tackled through iteratively solving a series of convex optimization problem \eqref{P22}. The objective value achieved by iteratively solving problem \eqref{P22} is non-decreasing over iterations. Moreover, since the optimal objective value is lower bounded, the convergence of the proposed algorithm for tackling problem \eqref{P21} is guaranteed theoretically after a sufficient number of iterations.

\subsection{IRS Phase Shift Design}

In this part, we optimize the IRS phase shift $\mathbf{v}_1$ and $\mathbf{v}_2$ under other given variables. By analyzing the corresponding problem, we obtain the following remark.

\begin{remark}
	Given the transmit covariance matrix $\mathbf{W}$, the $\mathbf{v}_2$  that maximizes the offloading rate $r^{\mathrm{c}}$ is always one of the optimal solutions. In the second stage, IRS assists the task offloading from ES to the CS. It can be observed that the design of $\mathbf{v}_2$ is only related to the transmit covariance matrix $\mathbf{W}$ in the ES and the CS computing capability. Note that we assume that CS is equipped with sufficient CPU computing capability, i.e., the offloaded task from the ES to the CS can always be successfully processed. Hence, the obtained $\mathbf{v}_2$ by maximizing the offloading rate $r^{\mathrm{c}}$ is always optimal.
\end{remark}

To obtain the corresponding solution presented in Remark 3, we first transform the offloading rate from ES to the CS into a more tractable form in terms of each entry $\{v_n\}$ in $\mathbf{v}_2$. Note that the corresponding composite channel is
$\tilde{\mathbf{H}}_{c} = \mathbf{H}_{c} + \mathbf{G}_c \mathbf{\Theta}_2 \mathbf{F}_{c}$.
Denote $\mathbf{G}_c = [\mathbf{g}_1, \ldots, \mathbf{g}_N]$ and $\mathbf{F}_{c} = [\mathbf{f}_1, \ldots, \mathbf{f}_N]^H$, where $\{\mathbf{g}_n\}$ is the columns of $\mathbf{G}_c$ and $\{\mathbf{f}_n^H\}$ is the rows of $\mathbf{F}_{c}$. Thus, the channel between the ES and the CS can be decoupled as
\begin{align}
	\tilde{\mathbf{H}}_{c} = \mathbf{H}_{c} + \sum_{n=1}^N v_n \mathbf{g}_n \mathbf{f}_n^H.
\end{align}
Besides, let $\mathbf{W} = \mathbf{U}_{W} \mathbf{\Sigma}_W \mathbf{U}_W^H $ denote the eigenvalue decomposition (EVD) of $\mathbf{W}$, and further define $\mathbf{H}_c' = \mathbf{H}_c \mathbf{U}_{W} \mathbf{\Sigma}_W^{\frac{1}{2}}$, $\mathbf{F}'_c = \mathbf{F}_c \mathbf{U}_{W} \mathbf{\Sigma}_W^{\frac{1}{2}} = [\mathbf{f}_1', \ldots, \mathbf{f}_N']^H$, where $\mathbf{f}_n' = \mathbf{\Sigma}_W^{\frac{1}{2}} \mathbf{U}_W^H \mathbf{f}_n$. 
Thus, we have \cite{ShuowenZhang_MIMO_IRS}
\begin{align}
	r^{\mathrm{c}}(\{v_n\}) = & \log _2 \operatorname{det}\left(\mathbf{I}_{M_2}+\frac{1}{\sigma^2} \mathbf{H}_c' \mathbf{H}'^H + \frac{1}{\sigma_{\mathrm{c}}^2} \sum_{i=1} \mathbf{g}_i \mathbf{f}_i'^H \mathbf{f}_i' \mathbf{g}_i^H\right. \nonumber \\
& +\frac{1}{\sigma_{\mathrm{c}}^2} \sum_{i=1}^N \sum_{j=1, j \neq i}^N v_i v_j^* \mathbf{g}_i \mathbf{f}_i^{\prime H} \mathbf{f}_j^{\prime} \mathbf{g}_j^H \\
& \left.+\frac{1}{\sigma_{\mathrm{c}}^2} \sum_{i=1}^N\left(\mathbf{H}_c^{\prime} v_i^* \mathbf{f}_i' \mathbf{g}_i^H + v_i \mathbf{g}_i \mathbf{f}_i'^H \mathbf{H}_c'^H\right)\right)
\end{align}
 Given $\{v_i\}, i \neq n $, $r_c$ can be reformulated as a explicit function of $v_n$ that is given by
\begin{align}
	r^{\mathrm{c}}(v_n) = & \log _2 \operatorname{det}\left(\mathbf{A}_n + v_n \mathbf{B}_n + v_n \mathbf{B}_n^H  \right),
\end{align}
where
\begin{align}
	&\!\!\!\!\!\!\!\mathbf{A}_n = \mathbf{I}_{M_2}+\frac{1}{\sigma_c^2}\left( \mathbf{H}_c' + \mathbf{G}_c \mathbf{\Theta}_2 \mathbf{F}_{c}' - v_n \mathbf{g}_n \mathbf{f}_n^{\prime H}\right) \nonumber \\
	& \times\left(\mathbf{H}_c' + \mathbf{G}_c \mathbf{\Theta}_2 \mathbf{F}_{c}' - v_n \mathbf{g}_n \mathbf{f}_n^{\prime H}\right)^H +\frac{1}{\sigma_c^2} \mathbf{g}_n \mathbf{f}_n^{\prime H} \mathbf{f}_n^{\prime} \mathbf{g}_n^H, \\
&\!\!\!\mathbf{B}_n = \frac{1}{\sigma_c^2} \mathbf{g}_n \mathbf{f}_n^{\prime H}\left(\mathbf{H}_c^{\prime H} + (\mathbf{G}_c \mathbf{\Theta}_2 \mathbf{F}_{c}' - v_n \mathbf{g}_n \mathbf{f}_n^{\prime H})^H\right).
\end{align}


Therefore, the subproblem with respect to each element $v_n$ in $\mathbf{v}_2$ can be expressed as
\begin{subequations} \label{Pv2}
	\begin{align}
		\mathop  {\max}\limits_{\scriptstyle v_n} \ \ &  \log _2 \operatorname{det}\left(\mathbf{A}_n + v_n \mathbf{B}_n + v_n \mathbf{B}_n^H  \right) \\
		{\rm{s.t.}}\ \ 
		& |v_n| = 1.
	\end{align}
\end{subequations}
By loosening the norm-one constraints to $|\overline{v}_n| \leq 1,\forall n$, we can obtain a convex optimization problem with respect to each element $v_n$. Then, we can subtract the obtained phase shift and construct $\mathbf{v}_2$.

After obtaining the IRS phase shift configuration $\mathbf{v}_2$ in $T_2$, the remaining variable is $\mathbf{v}_1$.
Similar to the proposed algorithm in Section \ref{sub1} for solving the subproblem \eqref{P2} related to the transceiver design and computation resource allocation, the corresponding problem is formulated as
\begin{subequations} \label{P3}
	\begin{align}
		\mathop  {\max}\limits_{\scriptstyle S_a, S_o,  
			\atop \scriptstyle I_a, \mathbf{v}_1 }   \ & w_a \mathrm{log}_2 ( \frac{P_{a}}{S_a I_a} )  + w_o \mathrm{log}_2 \big(1+ \frac{1}{S_o} \big) \label{} \\
		{\rm{s.t.}}\ \
		&|v_{1,n}| = 1, 1\leq n\leq N, \label{} \\
		&\eqref{P1_c_bits}, \eqref{P21_relax1}, \eqref{P21_relax2}, \eqref{P21_relax3}, \eqref{P22_Ra}.
	\end{align}
\end{subequations}
Note that similar to problem \eqref{P21}, the constraints in problem \eqref{P3} have been relaxed to inequalities without loss of optimality. 
In the subsequent, we first transform the composite channels induced by IRS to the more tractable forms which are given by
\begin{align}
	\mathbf{h}_{a,k} + \mathbf{G}^{H}_e \mathrm{diag}(\mathbf{f}_{a,k}) \mathbf{v}_1 &= [\mathbf{G}^{H}_e\mathrm{diag}(\mathbf{f}_{a,k}) \ \mathbf{h}_{a,k}] \times \begin{bmatrix}
		\mathbf{v}_1 \\
		1 
	\end{bmatrix} \nonumber \\
	& \triangleq \tilde{\mathbf{G}}^{H}_{a,k} \tilde{\mathbf{v}}_1, \\
	\mathbf{h}_{o,k} + \mathbf{G}^{H}_e \mathrm{diag}(\mathbf{f}_{o,k}) \mathbf{v}_1 &= [\mathbf{G}^{H}_e\mathrm{diag}(\mathbf{f}_{o,k}) \ \mathbf{h}_{o,k}] \times \begin{bmatrix}
		\mathbf{v}_1 \\
		1 
	\end{bmatrix} \nonumber \\
	& \triangleq \tilde{\mathbf{G}}^{H}_{o,k} \tilde{\mathbf{v}}_1.
\end{align}
We have
\begin{align} 
	&|\mathbf{m}^H \tilde{\mathbf{h}}_{a,k}|^2 = | \tilde{\mathbf{v}}_1^H \tilde{\mathbf{G}}_{a,k} \mathbf{m}|^2 = | \tilde{\mathbf{v}}_1^H \tilde{\mathbf{m}}_{a,k}|^2, \label{optv1_1} \\
	&\| \tilde{\mathbf{h}}_{o,k}\|^2 = \| \tilde{\mathbf{v}}_1^H \tilde{\mathbf{G}}_{o,k} \|^2. \label{optv1_2}
\end{align}
where $\tilde{\mathbf{m}}_{a,k} = \tilde{\mathbf{G}}_{a,k} \mathbf{m}$. 
Note that both \eqref{optv1_1} and \eqref{optv1_2} are convex functions with respect to $\tilde{\mathbf{v}}_1$. By applying the first-order Taylor expansion, for any given local point $\tilde{\mathbf{v}}_1^{(i)}$ in the $i$-th iteration, we have
\begin{align}
	\|\tilde{\mathbf{v}}_1^H  \tilde{\mathbf{m}}_{a,k}\|^2 \! &\geq \! - (\mathbf{v}_1^{(i)})^H \tilde{\mathbf{m}}_{a,k} \tilde{\mathbf{m}}_{a,k}^H  \mathbf{v}^{(i)}_1 \! + \! 2\Re{(\mathbf{v}_1^{(i)})^H \tilde{\mathbf{m}}_{a,k} \tilde{\mathbf{m}}_{a,k}^H  \mathbf{v}}_1,\nonumber \\
	&\triangleq f^{\mathrm{low}}_{a,k} (\mathbf{v}_1,\mathbf{v}_1^{(i)}). \\
	\| \tilde{\mathbf{v}}_1^H \tilde{\mathbf{G}}_{o,k} \|^2 \! &\geq \! - (\mathbf{v}_1^{(i)})^H \tilde{\mathbf{G}}_{o,k}\tilde{\mathbf{G}}_{o,k}^H \mathbf{v}^{(i)}_1 \! + \! 2\Re{(\mathbf{v}_1^{(i)})^H \tilde{\mathbf{G}}_{o,k}\tilde{\mathbf{G}}_{o,k}^H \mathbf{v}}_1,\nonumber \\
	&\triangleq f^{\mathrm{low}}_{o,k} (\mathbf{v}_1,\mathbf{v}_1^{(i)}).
\end{align}
Hence, regarding the constraints \eqref{P21_relax1} and \eqref{P21_relax3}, we utilize the SCA technique to transform them into
\begin{align}
	&\frac{1}{S_a} \leq f^{\mathrm{low}}_{a,k} (\mathbf{v}_1,\mathbf{v}_1^{(i)}), \label{P3_relax2} \\
	& \frac{1}{S_o} \leq \sum_{k\in \mathcal{K}_o}  \frac{ f^{\mathrm{low}}_{o,k} (\mathbf{v}_1,\mathbf{v}_1^{(i)}) }{ t_k \sigma^2_{e} }\label{P3_relax3}, 
\end{align}
where $\mathbf{v}_1^{(i)}$ is the given local point in $i$-th iteration. 

Besides, the objective function in \eqref{P3} is approximated as $w_a R^{\mathrm{low}}_a( S_a, I_a) + w_o R^{\mathrm{low}}_o (S_o) $ in each iteration. Hence, by loosening the norm-one constraints to $|\overline{v}_{i}| \leq 1,\ \forall i$, we can obtain a convex optimization problem with respect to $\mathbf{v}_1$ in each iteration that is given by
\begin{subequations} \label{P3_1}
	\begin{align}
		\mathop  {\max}\limits_{\scriptstyle S_a, S_o,  
			\atop \scriptstyle I_a, \mathbf{v}_1 }   \ & w_a \mathrm{log}_2 ( \frac{P_{a}}{S_a I_a} )  + w_o \mathrm{log}_2 \big(1+ \frac{1}{S_o} \big) \label{} \\
		{\rm{s.t.}}\ \
		& |v_{1,n}| \leq 1, 1\leq n\leq N, \\
		& v_{1,N+1} = 1, \\
		& \eqref{P1_c_bits}, \eqref{P21_relax2}, \eqref{P22_Ra},\eqref{P3_relax2},\eqref{P3_relax3}.
	\end{align}
\end{subequations}
By iteratively solving problem \eqref{P3_1} and reconstructing the obtained phase shifts as unit-modulus solutions, we can obtain the satisfying solution of $\mathbf{v}_1$.

\begin{remark}
	In practice, the IRS is generally implemented with discrete phase shifts. By considering that the phase-shift values are uniformly distributed in the interval $[0,2\pi)$, the corresponding feasible set is given by $\theta_{n} \in \mathcal{F} \triangleq\left\{0, \frac{2 \pi}{2^l}, \ldots, \frac{2 \pi}{2^l}\left(2^l-1\right)\right\}, \forall n \in \mathcal{N}$, where $l$ represents the number of quantization bits. After obtaining the beamforming vector $\mathbf{v}^*= [e^{\jmath \theta_1^*}, \ldots, e^{\jmath \theta_N}]^T$ via the former algorithms, the high-quality solution of the quantized phase shift can be obtained by $\theta'_{n} = \arg \min_{ \theta_{n}' \in \mathcal{F} } | \theta'_{n} - \theta_n^* |, \forall n \in \mathcal{N}$.
\end{remark}

\subsection{Time Allocation}

For any given transceiver design, phase shift configuration, and computation resource allocation, the time allocation optimization problem is given by
\begin{subequations} \label{Ptime}
	\begin{align}
		\mathop  {\max}\limits_{\scriptstyle T_1, T_2
			\atop \scriptstyle } \ \ \
		&R_{\mathrm{total}} \\
		{\rm{s.t.}}\ \ \ \;
		& \eqref{P1_c_time}, \eqref{P1_c_bits} .
	\end{align}
\end{subequations}
Note that all constraints and the objective function in problem \eqref{Ptime} are linear, problem \eqref{Ptime} is thus a convex optimization problem, which can be solved by the standard convex optimization techniques, such as the interior-point method.

\subsection{Overall Algorithm and Computational Complexity Analysis} \label{Complexity}

%
%

\begin{algorithm}[!t]
	\caption{Proposed Algorithm for solving  problem \eqref{P1}.}	\label{alg1}
	\begin{algorithmic}[1]
		\STATE  \textbf{Initialize} $\mathbf{m}^{(i)}, S_a^{(i)}, I_a^{(i)}, S_o^{(i)}, \{t_k^{(i)}\}, T_1, T_2, \mathbf{v}_1, \mathbf{v}_2$.
		\STATE  \textbf{repeat}
		\STATE  \quad Update $\mathbf{m}, \{t_k\}, \mathbf{W},
			 \{f_{k,1}^{\mathrm{lo}}, f_{k,2}^{\mathrm{lo}}\}, f^{\mathrm{es}}$ by solving problem \eqref{P22}.
		\STATE  \quad Update $\mathbf{v}_1, \mathbf{v}_2$ by solving problem \eqref{Pv2} and \eqref{P3_1}.  
		\STATE  \quad Update $T_1, T_2$ by solving problem \eqref{Ptime}. 	
		\STATE  \quad $i=i+1$. 
		\STATE \textbf{until}   The fractional increase of the objective function of \eqref{P1} is below a threshold $\varepsilon$.
		\STATE \textbf{return} The transmit power allocation of MEC UEs $ \{p_k\}$, receiver at ES $\mathbf{m}$, transmit signal covariance matrix at ES $\mathbf{W}$, IRS phase shift configuration $ \mathbf{v}_1, \mathbf{v}_2$, and the time allocation $T_1, T_2$.
	\end{algorithmic}
\end{algorithm}

Based on the three subproblems posed above, we propose an iterative algorithm for problem \eqref{P1} by utilizing the AO and SCA techniques, which is summarized in Algorithm 1. 
During the initial local point selection and the subsequent problem solving stages, the limited transmit power constraints on the AirComp UE poses a challenge as it may render the problem infeasible since we need the AirComp rate to be large than $0$. Consequently, it becomes necessary to first check whether the problem is feasible or not. It can be observed that the problem is feasible if
\begin{align} \label{feas_eq1}
	P_{a} \min_k |\mathbf{m}^{H} \tilde{\mathbf{h}}_{a,k}|^2 > \sum_{k \in \mathcal{K}_o} p_k |\mathbf{m}^H \tilde{\mathbf{h}}_{o,k} |^2 + \|\mathbf{m}\|^2\sigma_e^2.
\end{align}
It is worth noting that by assigning sufficiently low transmit power values $\{p_k\}$ to the MEC UE, the condition expressed in equation \eqref{feas_eq1} can be satisfied with a high probability.
Additionally, we can obtain the feasible local point by alternately optimizing $\mathbf{m}, \{p_k\}, \mathbf{v}_1$ until \eqref{feas_eq1} is satisfied. The corresponding construct problems are given by 
\begin{subequations} \label{feas1}
	\begin{align}
		\mathop  {\max}\limits_{\scriptstyle \mathbf{m}, \{t_k\}, \Gamma
			\atop \scriptstyle }   \ & 	P_{a} \Gamma - \sum_{k \in \mathcal{K}_o} |\mathbf{m}^H \tilde{\mathbf{h}}_{o,k} |^2/t_k - \|\mathbf{m}\|^2\sigma_e^2 \\
		{\rm{s.t.}}\ \
		&1/t_k \leq P_{o}, \forall k \in \mathcal{K}_{o}, \\
		&\Gamma \leq |\mathbf{m}^{H} \tilde{\mathbf{h}}_{a,k}|^2, \forall k \in \mathcal{K}_a, \label{feas1_m}
	\end{align}
\end{subequations}
\begin{subequations} \label{feas2}
	\begin{align}
		\mathop  {\max}\limits_{\scriptstyle \tilde{\mathbf{v}}_1, \{t_k\}, \Gamma
			\atop \scriptstyle }   \ & 	P_{a} \Gamma - \sum_{k \in \mathcal{K}_o} \| \tilde{\mathbf{v}}_1^H \tilde{\mathbf{G}}_{o,k} \mathbf{m} \|^2/t_k - \|\mathbf{m}\|^2\sigma_e^2 \\
		{\rm{s.t.}}\ \
		& 1/t_k \leq P_{o}, \forall k \in \mathcal{K}_{o}, \\
		& \Gamma \leq \| \tilde{\mathbf{v}}_1^H \tilde{\mathbf{m}}_{a,k}\|^2, \forall k \in \mathcal{K}_a, \label{feas1_v} \\
		& |v_{1,n}| \leq 1, 1\leq n\leq N, \\
		& v_{1,N+1} = 1.
	\end{align}
\end{subequations}
Finally, by transforming \eqref{feas1_m}, \eqref{feas1_v} into $\Gamma \leq f^{\mathrm{low}}_k (\mathbf{m},\mathbf{m}^{(i)})$, $\Gamma \leq f^{\mathrm{low}}_{a,k} (\mathbf{v}_1,\mathbf{v}_1^{(i)})$, respectively, and we then alternately solve problems \eqref{feas1} and \eqref{feas2} until \eqref{feas_eq1} becomes satisfied. We can obtain a feasible local point for the subsequent problem solving.

We then demonstrate the convergence of Algorithm $1$ and analyze its computational complexity.
In each iteration $i$, by optimally solving the subproblems at steps $3$, $4$, and $5$, we can obtain that  the corresponding objective functions' values satisfy
\begin{align} \label{p_analysis1}
	R_{\mathrm{total},3}^{\mathrm{low},(i)} \leq R_{\mathrm{total},4}^{\mathrm{low},(i)} \leq R_{\mathrm{total},5}^{(i)}.
\end{align}
Besides, due to the fact that the first-order Taylor expansions are tight at the given local points when solving problem \eqref{P22} and \eqref{P3}, it yields 
\begin{align} \label{p_analysis2}
	R_{\mathrm{total},3}^{\mathrm{low},(i)}  = R_{\mathrm{total},3}^{(i)},  R_{\mathrm{total},4}^{\mathrm{low},(i)}   = R_{\mathrm{total},4}^{(i)}.
\end{align}
Combing \eqref{p_analysis1} and \eqref{p_analysis2}, we can obtain that the lower bound value of the original objective function \eqref{P1_obj} is non-decreasing over iterations as well as the original objective function itself with optimally solving each subproblem in steps 3, 4, and 5. 
Besides, the objective value of \eqref{P1_obj} is upper bounded by a finite value due to the limited transmit power in UEs and ES, thus the proposed algorithm is guaranteed to converge to a locally optimal point.

The proposed algorithm for maximizing the hybrid computing rate of the considered DITEN system is shown as solving a series of convex problems.	
The main computational complexity of Algorithm 1 lies in steps 3, 4, and 5.
Specifically, problem \eqref{P22} in step 3 has four kinds of original variables, where the dimensions are given by $M_1$, $K_o$, $M_1^2$, $2K_o+1$. Besides the induced three auxiliary variables are scalars.  Hence, the corresponding computational complexity is given by ${\cal O}\big((M_1+M_1^2+3K_o+4)^{3.5}\big)$ via the interior-point method \cite{Interior_point_Journal}.
Similarly, the computational complexities of problem \eqref{Pv2} and \eqref{P3} in step 4 are ${\cal O}(N^{3.5})$, and  ${\cal O}\big((N+4)^{3.5}\big)$. 
Besides, problem \eqref{Ptime} in step 5 is shown as a linear programming problem, the corresponding computational complexity is given by ${\cal O}\big(2(N)^{2.5}\big)$. 
Hence, the overall computational complexity of the proposed algorithm is given by ${\cal O}\big(I_{\textrm{iter}}(M_1+M_1^2+3K_o+4)^{3.5} + (N+4)^{3.5}+N^{3.5}+2(N)^{2.5}) \big)$, in which $I_{\mathrm{iter}}$ represents the number of iterations required to achieve convergence.

\section{Simulation Results}

In this section, we provide numerical results to demonstrate the efficiency of the proposed algorithm for the considering DITEN system which comprises both multi-tier computing and environment monitoring capabilities. 
The ES, IRS, and CS are located at $(0,0,20)$ meter (m), $(0,2,20)$ m, $(-30,0,20)$ m, respectively. 
The placement of AirComp UEs is random within a radius of $5$ m that is centered at $(15, 15, 0)$ m, and the MEC UEs are randomly distributed within a circle centered at $(20, 20, 0)$ m with a radius $5$ m.
The distance-dependent path loss for all channels is modeled as $PL(d) = \rho_0 (d/d_0)^{-\alpha }$, where $\rho_0 = -30$ dB denotes the path loss at the reference distance $d_0=1$ m, $d$ denotes the link distance and $\alpha$ denotes the path loss exponent.
The path-loss exponents of the UEs-ES and ES-CS channels are set to 3.3, whereas the path-loss exponents for the UEs-IRS, ES-IRS, and IRS-CS channels are set to 2.3. 
Besides, the noise power at the ES and the CS are set as $\sigma_e^2 = \sigma_c^2 = -80$ dBm.
Unless otherwise stated, other parameters are set as: $B = 1$ MHz, $T=1$ s, $K_a = 20$, $K_o = 5$, $M_1=M_2=5$, $N=10$, $P_a = P_o = 5$ dBm, $P_{\mathrm{es}}$ = $20$ dBm, $\rho=500$ cycles/bit, $\kappa = 10^{-28}$.
The CPU frequency deviation of MEC UEs and the ES are given by $ \hat{f}_k^{\mathrm{lo}} = 0.1f_{\mathrm{lo},k}$, $ \hat{f}^{\mathrm{es}} = 0.1{f}^{\mathrm{es}}$, respectively. 
Furthermore, we consider the following schemes for comparison: 1) Hybrid computing system without the assistance of DT. 
Note that deviations in UEs and ES due to hardware impairments or other reasons are not available without the help of DT, thus the central controller cannot adapt to such mismatches. Specifically, we assume that the actually achieved computing frequency is ${f}_k^{\mathrm{lo}} -\hat{f}_k^{\mathrm{lo}} $ when the designer allocates $\kappa({f}_k^{\mathrm{lo}})^3$ power for task computing.
2) Offloading only strategy: all MEC UEs and the ES only perform the task ofﬂoading to higher tiers.
3) Single-tier computing system contains only the ES server.


\begin{figure}[t]  
	\centering
	\includegraphics[width=1.05\columnwidth]{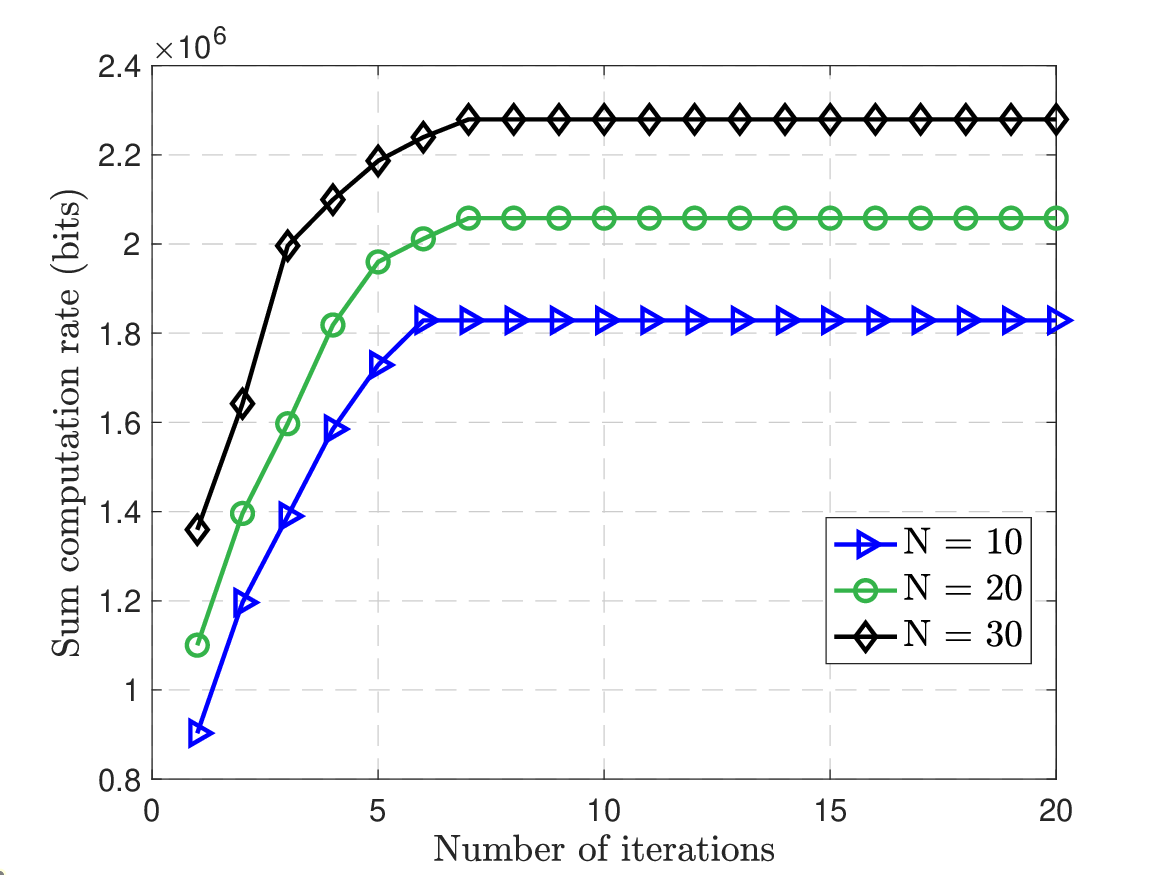} 
	\caption{Computation rate versus the number of iterations.}
	\label{convergence}
\end{figure}

In this section, we first present the convergence behavior of the proposed algorithm before analyzing the performance of the IRS-aided DITEN. The simulation result depicting the convergence behavior of the proposed algorithm is presented in Fig. \ref{convergence}. The figure clearly shows that the sum computation rate of the proposed algorithm increases rapidly with the number of iterations. 
It can be observed that the proposed algorithm converges after about $7$ iterations, which is consistent with the overall algorithm analysis presented in Section \ref{Complexity}. This result highlights the effectiveness of the proposed algorithm in addressing the NP-hard problem with highly coupled variables via AO and SCA techniques.

\begin{figure}[t]
	\centering
	\includegraphics[width=1.05\columnwidth]{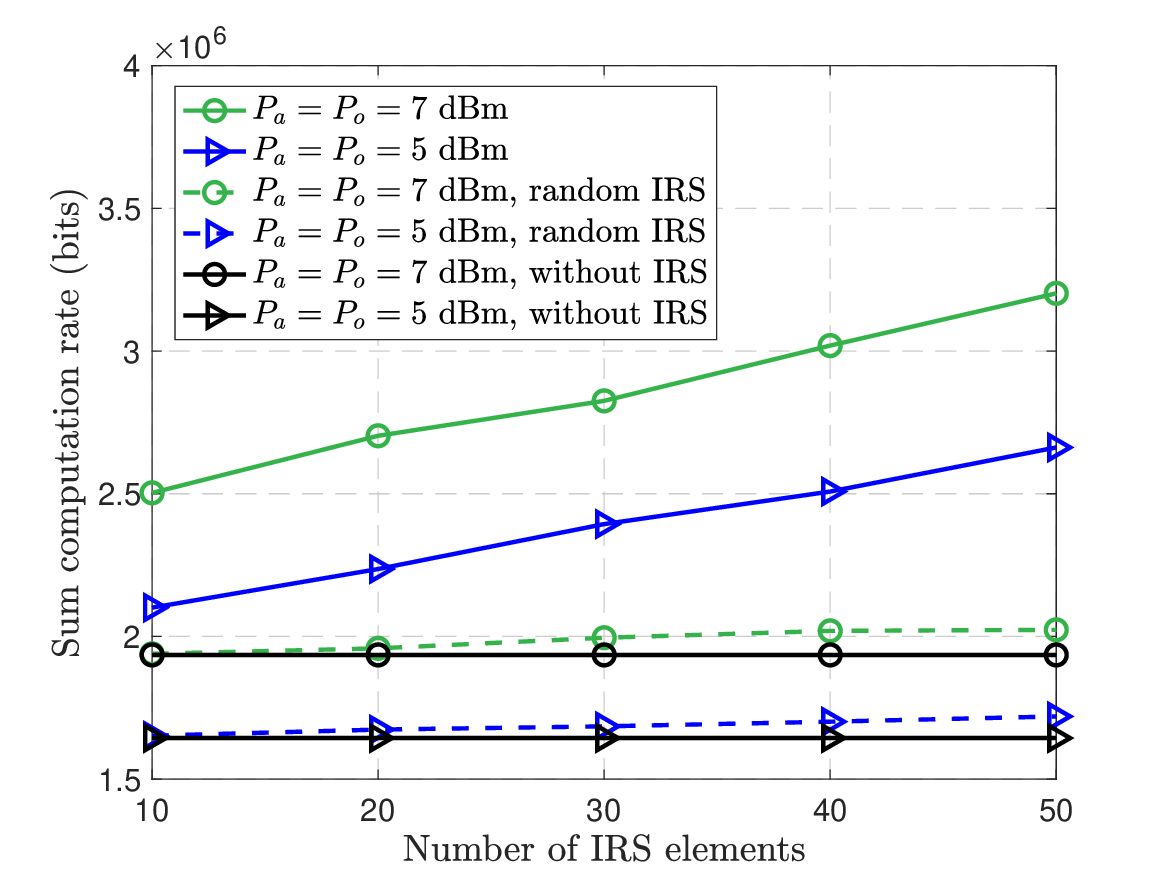}
	\caption{Computation rate versus the number of IRS elements.}
	\label{vsN}
\end{figure}

In Fig. \ref{vsN}, we compare the sum computation rate obtained by all schemes versus the number of IRS elements $N$.
One can see that the achieved performance improvement progressively with the increase of elements number. This is due to the fact that a larger number of reflecting elements help to achieve a higher passive beamforming gain, which is beneficial for signal transmission in different tiers. Moreover, it is observed that the proposed algorithms outperform the cases where phase shifts of the IRS are randomly given or without the assistance of the IRS. The performance gap between the proposed algorithms and the other cases widens as the number of IRS elements $N$ increases. This is because the proposed algorithms exploit the IRS to its full potential by intelligently selecting the phase shifts of the reflecting elements to maximize the sum computing rate. This observation demonstrates that with the aid of the IRS, it is possible to achieve a significant improvement in the sum computation rate. This improvement becomes more pronounced as the number of reflecting elements in the IRS increases.

\begin{figure}[t]
	\centering
	\includegraphics[width=1.05\columnwidth]{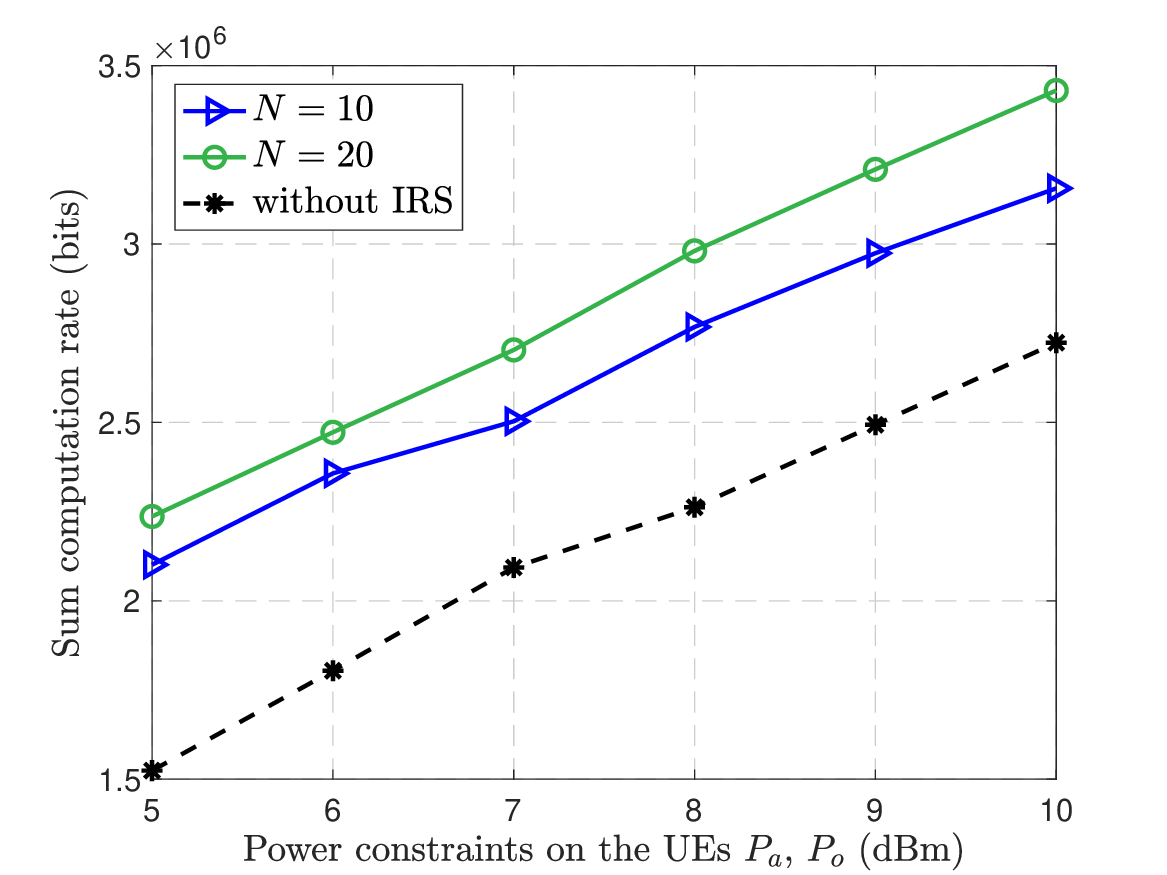}
	\caption{Computation rate versus the power constraints of UEs.}
	\label{vs_power}
\end{figure}

\begin{figure}[t]
	\centering
	\includegraphics[width=1.05\columnwidth]{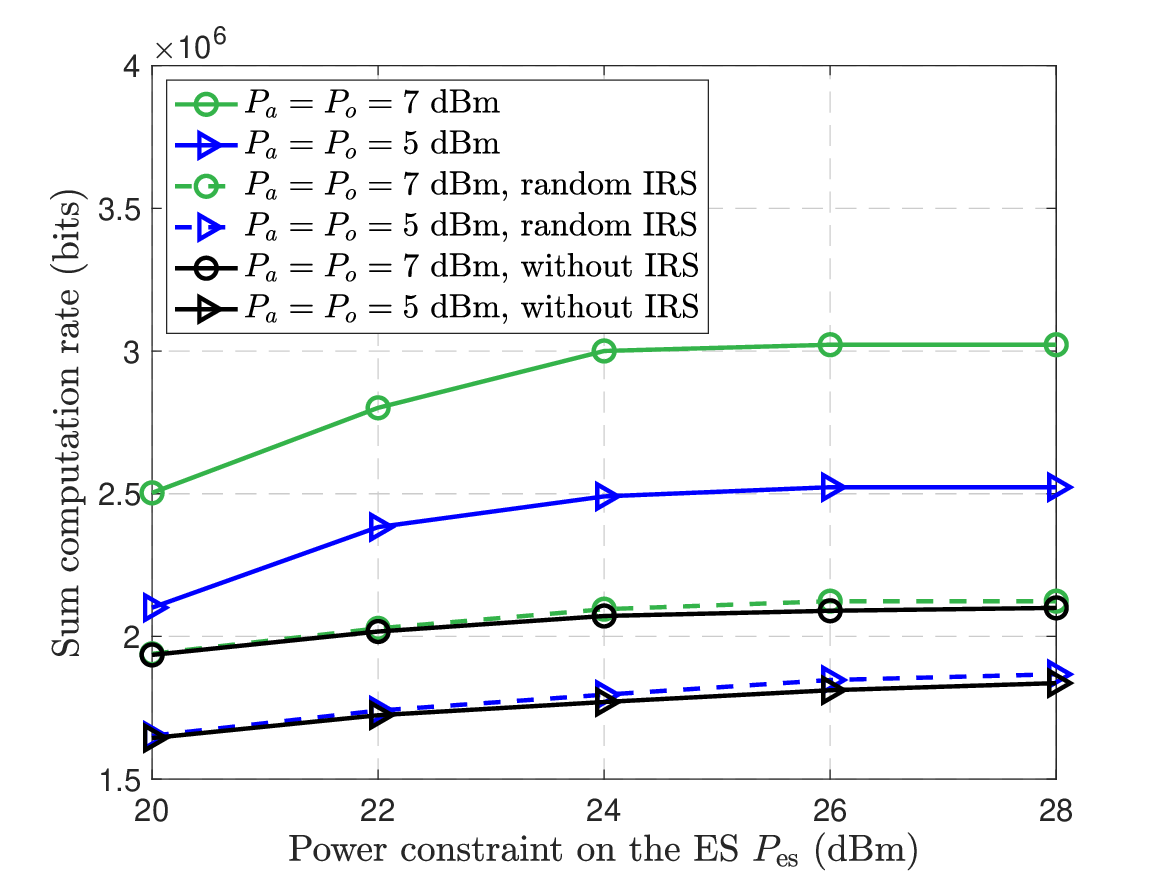}
	\caption{Computation rate versus the power constraint of the ES.}
	\label{vs_powerES}
\end{figure}
Fig. \ref{vs_power} shows the sum computation rate under various power constraints on UEs. It can be observed that by enlarging the available power of UEs, the achieved sum computation rate is monotonically increased. 
By comparing Fig. \ref{vs_power} to Fig. \ref{vsN}, we can observe that the IRS with a large number of elements achieves the same performance compared to the UE with larger power. The results demonstrate that increasing the number of elements in the IRS allows for substantial power savings at the UEs that can achieve higher energy efficiency. 
Furthermore, we present simulation results of the sum computation rate versus the power constraint of the ES in Fig. \ref{vs_powerES}. 
It is observed that a higher computation rate can be achieved with the increase of power constraint of the ES since more communication and computation resources are available for task computing and offloading. However, it is noticed that the achieved computation rate of all the schemes increase progressively reaches saturation, and finally bounded by a certain value. This is due to that either the task computing and offloading rate at the ES are limited by the task offloading rate from the MEC UEs to the ES.
Combine Fig. \ref{vs_power} and Fig. \ref{vs_powerES}, we can observe that the improvement on the computation rate of the multi-tier computing system requires precise coordination of the available resources in different tiers, particularly in the first tier corresponding to the original task generator. This is because the limitations in the first tier can significantly impact the overall system performance.

\begin{figure}[t]
	\centering
	\includegraphics[width=1.05\columnwidth]{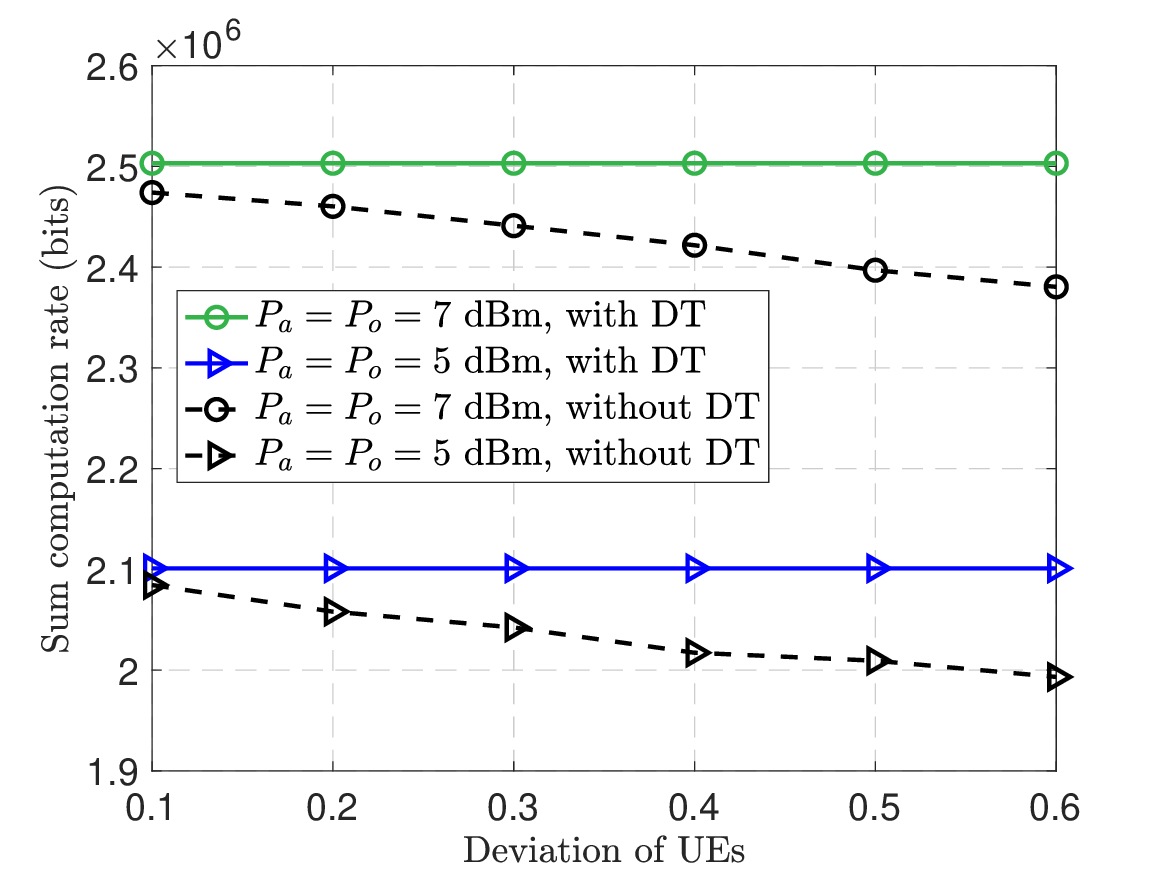}
	\caption{Computation rate versus the deviation of UEs.}
	\label{vsDevUE}
\end{figure}

\begin{figure}[t]
	\centering
	\includegraphics[width=1.05\columnwidth]{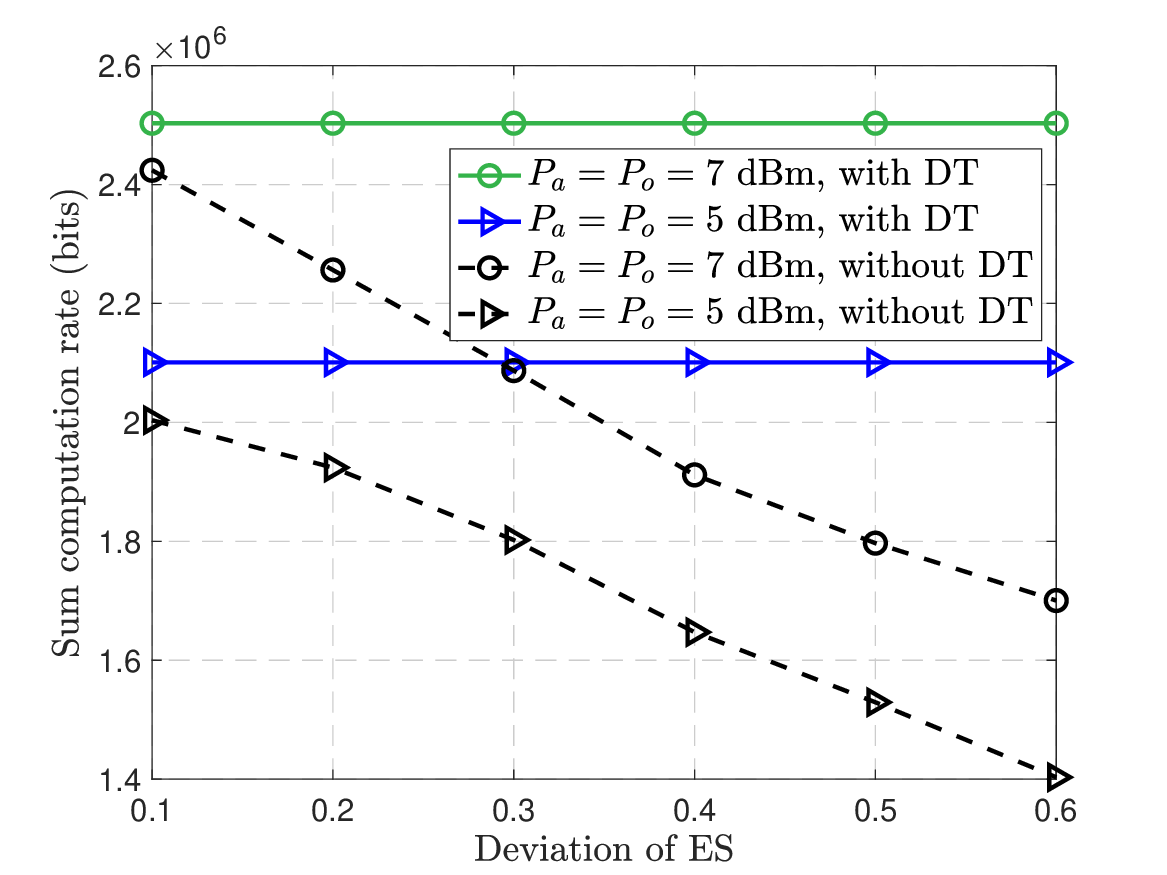}
	\caption{Computation rate versus the deviation of ES.}
	\label{vsDevES}
\end{figure}

To demonstrate the benefits of DT in improving the performance of DITEN, we compare the achieved computing rate of DITEN to the one without the assistance of DT.
Fig. \ref{vsDevUE} presents the performance versus the deviation of the MEC UEs. It can be observed that the achieved sum computation rate of the DITEN system remains unchanged since the DT system can precisely obtain the real-time operation status of the nodes. However, the sum computation rate of the system without DT monotonic decreases when the deviation becomes large. 
This occurs because the unknown deviation results in inaccurate design, thereby negatively impacting the system's performance.
As the deviation increases, the system design becomes increasingly inaccurate, which leads to degraded performance and reduced efficiency. However, the integration of DT into the DITEN system enables it to adapt to deviations and maintain optimal performance, thereby ensuring stable and consistent computation rates. 
Fig. \ref{vsDevES} depicts the relationship between the sum computation rate and the deviation of ES. It is noted that the corresponding performance is still monotonic decreases with the increase of deviation on ES. Additionally, the sum computation rate decays more quickly than that in Fig. \ref{vsDevUE}. This is primarily due to two reasons. Firstly, the computation resources available at the UEs are more limited compared to the ES, resulting in a smaller absolute deviation. Secondly, the UEs typically offload their tasks to the ES for faster computing, which means that the majority of the computational workload is handled by the ES rather than the UEs. Therefore, the deviation at the ES has a greater influence on system performance than that at the UEs. 
Combine Fig. \ref{vsDevUE} and Fig. \ref{vsDevES}, it can be observed that the integration of DT in the MEC system is crucial for ensuring its optimal performance by precisely allocating the available resources.

\begin{figure}[t]
	\centering
	\includegraphics[width=1.05\columnwidth]{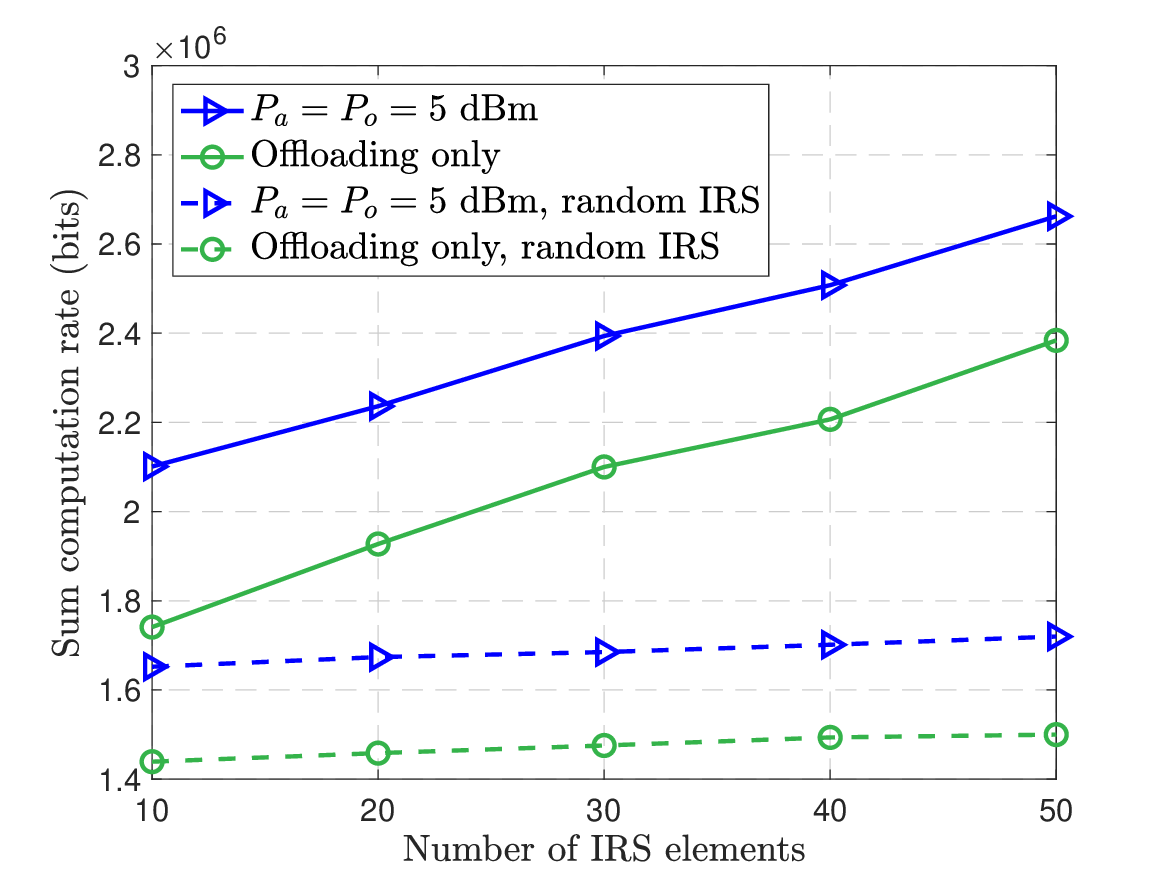}
	\caption{Computation rate compared to the offloading only strategy.}
	\label{vs_offloading}
\end{figure}

\begin{figure}[t]
	\centering
	\includegraphics[width=1.05\columnwidth]{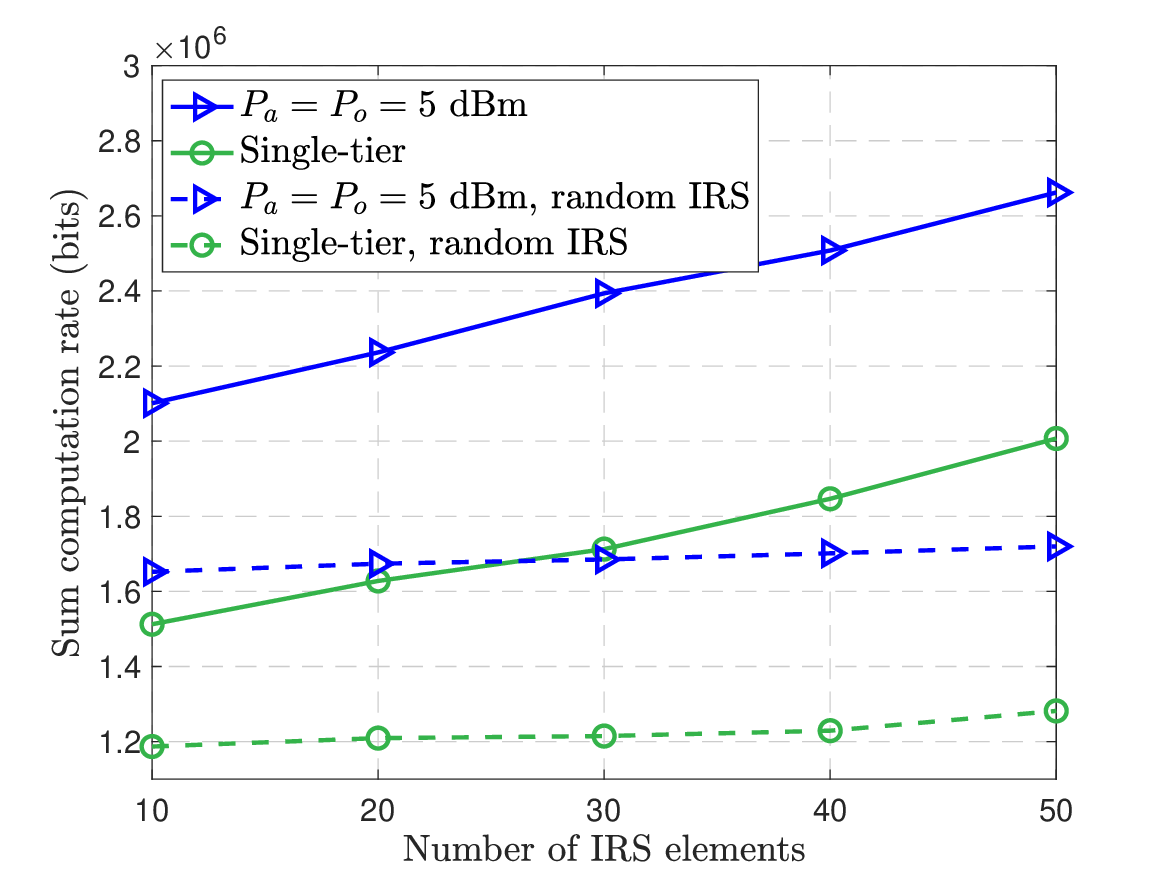}
	\caption{Computation rate compared to the single-tier computing system.}
	\label{singletier}
\end{figure}

In the subsequent, we compare the achieved computation rate of the proposed algorithm with the offloading only strategy. From Fig. \ref{vs_offloading}, the obtained offloading strategy outperforms the offloading only scheme, demonstrating the effectiveness of the proposed algorithm in optimizing the task offloading strategy. This result highlights the benefits of jointly determining offloading strategy and resource allocation in the multi-tier computing system architecture.
Furthermore, Fig. \ref{singletier} shows the performance comparison with a single-tier computing system. 
The multi-tier network architecture attains significantly higher total computation rate due to its ability to flexibly coordinate communication and computation resources across different tiers. In contrast, the single-tier system lacks this flexibility and is more constrained in resource coordination.
In summary, these results validate the advantages of the proposed algorithm and multi-tier computing framework. By intelligently mapping tasks and allocating resources, significant performance gains over both static offloading and conventional single-tier systems can be achieved.

\section{Conclusion}

In this paper, we have proposed a novel multi-tier hybrid computing framework assisted by an IRS for the DITEN system.
The proposed framework offered the opportunity to construct an integrated system for DT and MEC that aimed to achieve the network-wide convergence.
Specifically, AirComp was utilized to perform efficient system monitoring to construct the DT layer, while DT assisted the MEC to achieve precise task offloading. In the considered system, an IRS was employed to enhance the signal transmission among different tiers and suppress the interference among heterogeneous nodes that facilitated efficient and reliable communication. 
We proposed an iterative algorithm based on the AO and SCA techniques to handle the resulted NP-hard problem in such a complex system.
Simulation results have demonstrated the efficiency and effectiveness of the proposed algorithm. In particular, a significant performance gain can be achieved with the assistance of the IRS.
Furthermore, the results indicated that the balance between local computing and task offloading can be precisely achieved in the DITEN system, as real-time system status can be obtained witxh the help of DT. This paper highlighted the importance of DT in the integration of DT and MEC and demonstrated its necessity in achieving an optimal performance in DITEN systems through analysis and numerical results.

\balance

\bibliographystyle{IEEEtran}
\bibliography{refs}

\end{document}